\documentclass[a4paper,fleqn]{cas-sc}

\usepackage[numbers,longnamesfirst]{natbib}

\usepackage[english]{babel}
\usepackage{booktabs}
\usepackage{colortbl}
\usepackage{comment}
\usepackage{graphicx}
\usepackage{epsfig}
\usepackage{array, colortbl}
\usepackage{listings}
\usepackage{epstopdf}
\usepackage{multirow}
\usepackage{rotating}
\usepackage{xspace}
\usepackage{float}
\usepackage{amsmath}

\usepackage{balance}
\usepackage{fancybox}
\usepackage{scalefnt}
\usepackage{amssymb}
\usepackage[normalem]{ulem}
\usepackage{mathtools}

\usepackage{enumitem}

\usepackage{tabularx,multirow,blindtext}
\usepackage{graphicx}
\usepackage{subcaption}
\usepackage{caption}
\usepackage[skins]{tcolorbox}
\usepackage{array}
\newcolumntype{C}[1]{>{\centering\arraybackslash}p{#1}}

\usepackage{makecell}

\usepackage{ragged2e}

\makeatletter
\renewcommand{\paragraph}[1]{\noindent\textsf{#1}.}

\definecolor{britishracinggreen}{rgb}{0.0, 0.26, 0.15}
\definecolor{cadmiumorange}{rgb}{0.93, 0.53, 0.18}
\definecolor{darkbyzantium}{rgb}{0.36, 0.22, 0.33}
\definecolor{darkgreen}{rgb}{0.0, 0.2, 0.13}
\definecolor{oxfordblue}{rgb}{0.0, 0.13, 0.28}
\definecolor{palatinateblue}{rgb}{0.15, 0.23, 0.89}
\definecolor{pansypurple}{rgb}{0.47, 0.09, 0.29}

\usepackage{cleveref}

\usepackage{url}

\newcommand{\conclusion}[1]{\begin{center}\begin{tcolorbox}[skin=widget, left=2mm,right=2mm,top=2mm,bottom=2mm,boxrule=0.3mm,arc=0mm,coltitle=black,colframe=black!99!white,colback=white!99!gray,width=(\linewidth),before=\hfill,after=\hfill]#1\end{tcolorbox}\end{center}}

\newcommand*{\rom}[1]{\expandafter\@slowromancap\romannumeral #1@}

\newcommand{\npm}{\textit{\textit{npm}}\xspace}
\newcommand{\sv}{{{semantic versioning}}\xspace}

\newcommand{\rqi}{RQ1: Can we effectively determine the \sv type of a new package release?}
\newcommand{\rqii}{RQ2: Which dimension of features are most important in determining the \sv type of a new package release?}
\newcommand{\rqiii}{RQ3: How effective are the machine learning techniques when applied on cross-packages?}

\begin{document}
	
	
	\shortauthors{Abdalkareem  et~al.}
	
	\title [mode = title]{A Machine Learning Approach to Determine the Semantic Versioning Type of \npm Packages Releases}                      
	
	

	%
	
	\author[1]{Rabe Abdalkareem}[orcid=0000-0001-9914-5434]
	\ead{rabe.abdalkareem@carleton.ca}
	\affiliation[1]{organization={School of Computer Science, Carleton University},city={Ottawa},country={Canada}}

	\author[2]{Md Atique Reza Chowdhury}
	\ead{mwdhu@encs.concordia.ca}
	\affiliation[2]{organization={Data-driven Analysis of Software (DAS) Lab, Concordia  University},city={Montreal},country={Canada}}

	\author[2]{Emad Shihab}[orcid=0000-0003-1285-9878]
	\ead{eshihab@encs.concordia.ca}

\begin{abstract}
Semantic versioning policy is widely used to indicate the level of changes in a package release. Unfortunately, there are many cases where developers do not respect the semantic versioning policy, leading to the breakage of dependent applications. To reduce such cases, we proposed using machine learning (ML) techniques to effectively predict the new release type, i.e., patch, minor, major, in order to properly determine the semantic versioning type. To perform our prediction, we mined and used a number of features about a release, such as the complexity of the changed code, change types, and development activities. We then used four ML classifiers. To evaluate the performance of the proposed ML classifiers, we conducted an empirical study on {31} JavaScript packages containing a total of approximately {6,260} releases. We started by extracting 41 release-level features from historical data of packages' source code and repositories. Then, we used four machine learning classifiers, namely XGBoost, Random Forest, Decision Tree, and Logistic Regression. We found that the XGBoost classifiers performed the best achieving median ROC-AUC values of 0.78, 0.69, and 0.74 for major, minor, and patch releases, respectively. We also found that features related to the change types in a release are the best predictors group of features in determining the \sv type. Finally, we studied the generalizability of determining the \sv type by applying a cross-package validation. Our results showed that the general classifier achieved median ROC-AUC values of 0.76, 0.69, and 0.75 for major, minor, and patch releases. 
\end{abstract}
	
	
	
\begin{keywords}
	\npm Package Releases \sep Semantic Version \sep Mining Software Repository \sep Machine Learning	
\end{keywords}

    	\maketitle

	\section{Introduction}
	\label{sec:introduction}
	Semantic versioning is a commonly used versioning approach to signal a change's compatibility through version numbers. Prior work showed that properly adapting \sv increases developers' trust in their dependent on packages and decreases the chance of facing backward compatibility breakage~\cite{raemaekers2017semantic,BogartFSE2016}. Therefore, most language-specific package managers encourage the use of \sv (e.g., \npm for JavaScript, Cargo for Rust, Gems for Ruby, among others)~\cite{DecanTSE2019,DecanEMSE2019}. Likewise, some of the biggest software producers such as Microsoft, Netflix, Facebook, and Google significantly use \sv to tag their new software releases~\cite{LauingerDependency,potvin2016google,YarnAnew90online}. In addition, a survey with two thousand developers shows that developers heavily rely on \sv to determine the version of their projects' release type ~\cite{HowEcosy57online}.

However, misuse of \sv can cause many problems. Developers may incorrectly identify the \sv type and may tag a new release as minor or patch even though it introduces breaking changes, especially for packages that are continuously releasing~\cite{BogartFSE2016,857922fo45online}. One example of such a problem is in the context of the web browser Firefox and the font selection library fontconfig~\cite{857922fo45online}. At some point, the fontconfig's developers decided to change its implementation so that blank file names would no longer be permitted. They chose to mark this change as a minor release. However, this release of fontconfig caused Firefox to fail to render text for any application that used that minor release. In addition, this issue of release tagging can be particularly problematic for oversized packages or projects that receive many contributions and perform many changes in one release development duration. Therefor, this problem can negatively affect both the developers of the packages and software applications that directly or indirectly depend on these packages~\cite{BogartFSE2016,raemaekers2017semantic}.

Due to the increased adoption of \sv, most of the previous work focused on empirically studying its usage and benefits (e.g,. \cite{BogartFSE2016,Kula_EMSE2017,WitternMSR2016}). However, very few studies tried to improve the efficiency of applying the \sv in practice. More importantly, most of the prior studies took reactive approaches and tried to detect breakage changes of a package after it was released through the use of source code analysis (e,g., \cite{mujahidusingMSR2020,raemaekers2017semantic,MostafaISSTA2017,XavierSANER2017}). Thus, we argue that prior approaches have two key limitations. First, they tackled the issue of wrongly tagged releases after they are out and being integrated by others depending on applications. Second, they heavily relied on source code analysis, which suffers from high false-positive rates and is incapable of detecting runtime changes, especially for packages that are written in dynamic type language such as JavaScript~\cite{pradel2015typedevil,AndreasenACM2017}.

Therefore, the main goal of our work is to automatically determine the type of the new package release, i.e., patch, minor, and major. To do so, we proposed the use of machine learning (ML) techniques to predict the \sv type. 
We started by analyzing the \npm package manager and selected 31 packages with {6,268} releases that their developers properly use \sv to tag their releases. We then analyzed the source code and mined the development history of the studied packages, and extracted {41} features that are grouped into six dimensions, namely, change types, development activities, complexity and code, time, dependency, and text dimensions. Next, we built four different machine learning classifiers, namely XGBoost, Random Forest, Decision Tree, and Logistic Regression, to determine the \sv type of the releases. Finally, to evaluate the effectiveness of using the ML techniques, we performed an empirical study to answer the following questions:

\textit{\rqi}
We built four different ML classifiers using 41 features extracted from packages' repositories and source code. We then compared their performance to the baseline, which is the ZeroR classifier. Our results showed that XGBoost classifiers achieved average ROC-AUC values of {0.77, 0.69, and 0.74} (median $=$ 0.78, 0.69, and 0.74) for major, minor, and patch releases, respectively. In addition, this improvement equates to an average improvement of 1.58$X$, 1.38$X$, and 1.49$X$ by the built classifiers when they were compared to our baseline for the major, minor, and patch releases.


Then, we examined the most important dimension of features used by the ML classifiers to determine the \sv type of a new package release in order to provide insights to practitioners as to what features best indicate the new package release type. This led us to ask the question;
\textit{\rqii} 
We built different classifiers based on each dimension of features and evaluated and compared their performance. Our results showed that change types (e,g., number of JavaScript files added in a release.) and complexity of the source code of the release are the most important dimension of features in determining the type of new release.  
 
Lastly, to examine the generalizability of the proposed technique, we investigated the effectiveness of the ML techniques in determining the \sv type of a new package release using cross-packages validation. In particular, we asked the question; \textit{\rqiii} 
We built general classifiers and evaluated their performance using cross-package validation. The results showed that the classifier achieves average ROC-AUC values of 0.74, 0.68, and 0.75 (median $=$ 0.76, 0.69, and  0.75) for major, minor, and patch releases. These results also showed that cross-package classifiers' performances correspond to an average ROC-AUC improvement of 1.5$X$, 1.4$X$, and 1.5$X$ over our baseline. \newline

\noindent In general, our work made the following key contributions:
\begin{enumerate}
\item We formulated the problem of predicting \sv for JavaScript packages. To the best of our knowledge, this is the first work of using ML techniques to determine \sv type for JavaScript packages. We envision that our approach can be used to predict the releases that are likely to be breakage releases.

\item We proposed features that can be mined from JavaScript package repositories and source code to predict \sv type of a new package release. We used the proposed features to predict \sv accurately and studied the features that best indicate the \sv type.

\item We performed an empirical study on {31} open-source JavaScript packages, and our experimental results showed that the use of ML techniques can achieve an improvement over our baseline approach, which is the ZeroR classifier.

\end{enumerate}
\textbf{Structure of the paper:}
The remainder of the paper was organized as follows. Section~\ref{sec:SemanticVersioning} provided a background on \sv.
We described our case study design in Section~\ref{sec:approach}. We presented our case study results in Section~\ref{sec:case_study_results}. 
The work related to our study was discussed in Section~\ref{sec:related_work} and the threats to validity of our work is discussed in Section~\ref{sec:threats_to_validity}. Finally, Section~\ref{sec:conclusion} concluded the paper.


	\section{Semantic Versioning}
	\label{sec:SemanticVersioning}

Since the primary goal of our work is to determine the \sv type of a new \npm package release, it is essential first to provide background on the concept of \sv and how it is used to tag new package releases.

Semantic Versioning is considered the de-facto versioning standard for many software ecosystems, including node package manager (\npm) and Python package index (PyPI), to name a few. Semantic Versioning was introduced by the co-founder of GitHub, Tom Preston-Werner, in 2011. In our study, we focused on \sv 2.0, which was released in 2013 \cite{semver2019}. The purpose of \sv is twofold. It first allows package developers to communicate the extent of backward-incompatible changes in their new releases to application dependents. Also, it allows for dependents of a package to specify how restrictive or permissive they want to be in automatically accepting new versions of the packages.

In general, \sv proposes three dot-separated numbers indicating the major, minor, and patch versions of a release. Those numbers assist in identifying the type of changes in the newly released package. 
To explain how \sv works, we take the release \texttt{m1.n1.p1} number as an example. The first part \texttt{m1} presents the major type, the number \texttt{n1} stands for the minor type, and the number \texttt{p1} stands for the patch type. The \sv also shows rules for developers to determine how one of the three types number should be incremented when a new release comes out. In particular, any change to the new release package that is backward-incompatible (e.g., break the API) requires an update to the {major} version. 
Thus, a major release must yield the increment of the major version type, for example, from \texttt{m1.n1.p1} to \texttt{m2.n1.p1}. A minor release should be published when some new backward-compatible change is introduced (e.g., adding or supporting new functionality that does not create backward incompatibility). A minor release must yield the increment of the minor type of the version number (e.g., from \texttt{m2.n1.p1} to \texttt{m2.n2.p1}). Finally, a patch release should be published when the release represents backward compatible fixes (e.g., fixing a bug). A patch release must yield the increment of the patch type of the version number, such as from \texttt{m2.n2.p1} to \texttt{m2.n2.p2}. In addition, there are some optional {tags} for example specifying pre-releases type (e.g., \texttt{1.2.3}-beta).


Although adopting the semantic version is not mandatory, prior studies showed that mainly packages in \npm comply with this specification~(e.g., \cite{DecanTSE2019,AbbasTSE2021}). The mechanism to resolve a provided version relies on the precedence between version numbers since \npm needs to know if a particular version number is greater than, less than, or equal to another version number. Similar to decimal numbers, semantic version numbers are compared initially by the magnitude of their major type, then by their minor and finally by patch types. For example, version \texttt{3.2.1} is lower than versions \texttt{4.0.0} (by a major), \texttt{3.3.1} (by a minor), and \texttt{3.2.2} (by a patch), but greater than versions \texttt{2.2.1} (by a major), \texttt{3.1.1} (by a minor), and \texttt{3.2.0} (by a patch).

While \sv is a promising technique to specify the type of changes in a new package release, and even though it is recommended by ecosystem maintainers~\cite{AboutsSVonline}, it is not always straightforward to be used in practice. For example, a package developer can mistakenly flag the new release as a patch release while it is actually a major release. Therefore, this mistake might lead to many problems, mainly breaking the applications that depend on this package. In this paper, we formulated the determination of \sv type of a new package release as a research problem, which aimed to facilitate \npm packages developers to find the right \sv type for their new release packages. As a result, this will increase the packages' trust and reduce the breaking of applications that depend on those packages. 

	\section{Case Study Design}
	\label{sec:approach}


\begin{table}[t]
	\caption{The selection steps of the studied JavaScript packages that are published on \npm.}
	\centering
		\label{tab:selection_step}
		
	\begin{tabular}{l|r}
\hline
		\textbf{Selection Step} & \textbf{\# Packages}  \\ \hline
		\hline
		Most starred packages& 100 \\
		Packages without post- and pre- releases& 96 \\
		Packages with more than 50 releases& 77 \\
		Packages without breakage releases& 36 \\ \hline 
	\end{tabular}
\end{table}

The main {goal} of our study is to automatically determine the \sv type of a new release of a JavaScript package. To achieve this goal, we proposed the use of machine learning techniques. We begin by selecting JavaScript packages with a sufficient number of releases, and their developers use \sv to identify the type of the new releases. Next, we used the selected \npm packages as a labelled dataset. Then, we mined the source code and development history of the selected JavaScript packages to extract release-level features and used them as dependent variables in our machine learning classifiers. In the following subsections, we detail our labelled dataset, data extraction and processing steps, and the training of our classifiers.

\begin{table*}[bth]
	\centering
	\caption{Statistics of the studied JavaScript packages. The Table shows the name, number of commits, releases, analyzed releases, percentage of major, minor, patch releases of the studied packages.}
	\label{tab_studied_packages}
	  \scalebox{0.9}{
\begin{tabular}{p{1.4in}|r|r|r|r|r|r}
	\hline
	\textbf{Package} & \textbf{~Commits} & \textbf{~Release} & \textbf{~Analyzed} & \textbf{~\%Major} & \textbf{~\%Minor} & \textbf{~\%Patch} \\ \hline \hline

	renovate & 5,226 & 2293 & 1156 & 0.61 & 23.44 & 75.95 \\
	turtle.io & 1,110 & 413 & 294 & 2.38 & 8.16 & 89.46 \\
	sweetalert2 & 1,924 & 327 & 266 & 2.63 & 20.68 & 76.69 \\
	seek-style-guide & 579 & 280 & 222 & 10.81 & 39.19 & 50.00 \\
	oui & 722 & 226 & 207 & 4.35 & 5.31 & 90.34 \\
	react-isomorphic-render & 977 & 286 & 176 & 5.68 & 6.82 & 87.50 \\
	reactive-di & 625 & 133 & 107 & 6.54 & 8.41 & 85.05 \\
	module-deps & 492 & 135 & 104 & 5.77 & 30.77 & 63.46 \\
	express-processimage & 595 & 122 & 102 & 7.84 & 39.22 & 52.94 \\
	sku & 340 & 122 & 101 & 5.94 & 31.68 & 62.38 \\
	bittorrent-dht & 633 & 115 & 97 & 8.25 & 38.14 & 53.61 \\
	nightwatch-cucumber & 634 & 132 & 97 & 9.28 & 21.65 & 69.07 \\
	socketcluster-server & 282 & 111 & 94 & 12.77 & 27.66 & 59.57 \\
	eslint-config-canonical & 360 & 133 & 90 & 14.44 & 22.22 & 63.33 \\
	patchbay & 2,031 & 108 & 87 & 6.90 & 43.68 & 49.43 \\
	penseur & 210 & 95 & 81 & 8.64 & 50.62 & 40.74 \\
	mongo-sql & 511 & 87 & 78 & 7.69 & 12.82 & 79.49 \\
	pacote & 615 & 102 & 77 & 10.39 & 20.78 & 68.83 \\
	octokit/routes & 645 & 99 & 77 & 15.58 & 29.87 & 54.55 \\
	box-ui-elements & 1,329 & 88 & 72 & 9.72 & 52.78 & 37.50 \\
	rtc-quickconnect & 661 & 92 & 72 & 9.72 & 47.22 & 43.06 \\
	terrestris/react-geo & 2,846 & 73 & 69 & 11.59 & 46.38 & 42.03 \\
	rtcpeerconnection & 311 & 82 & 67 & 8.96 & 26.87 & 64.18 \\
	speakingurl & 429 & 78 & 66 & 19.70 & 28.79 & 51.52 \\
	license-checker & 377 & 70 & 65 & 35.38 & 18.46 & 46.15 \\
	octokit/fixtures & 378 & 81 & 64 & 12.50 & 51.56 & 35.94 \\
	repofs & 574 & 73 & 63 & 11.11 & 23.81 & 65.08 \\
	jsonrpc-bidirectional & 511 & 97 & 62 & 11.29 & 40.32 & 48.39 \\
	nes & 370 & 67 & 61 & 14.75 & 34.43 & 50.82 \\
	zapier-platform-cli & 1,003 & 69 & 61 & 11.48 & 27.87 & 60.66 \\
	rtc-signaller & 546 & 79 & 60 & 10.00 & 41.67 & 48.33 \\ \hline
	\textbf{Mean} & \textbf{898.30} & \textbf{202.20} & \textbf{138.50} & \textbf{10.09} & \textbf{29.72} & \textbf{60.20} \\
	\textbf{Median} & \textbf{595.00} & \textbf{102.00} & \textbf{81.00} & \textbf{9.72} & \textbf{28.79} & \textbf{59.57} \\ \hline
\end{tabular}

}

\end{table*}


\subsection{Test Dataset}
\label{TestDatset}
To perform our study, we needed to obtain a number of JavaScript packages that follow \sv guidelines to mark their releases type. To build our labelled dataset, we started by looking at JavaScript packages that are published on the Node Package Manager (\npm). We chose \npm package manager as it is the official registry and repository for JavaScript packages.

To collect our dataset, we resorted to the public repository of \npm that contains a list of all the published packages on \npm~\cite{npmregis30online}. The \npm repository contains metadata about every published package, such as the different releases of a package, the date of each release, and the release type. Since there are a large numbers of packages published on \npm and some of them did not provide high-quality packages~\cite{AbdalkareemFSE2017}, we had to apply filtration steps to select the packages that we wanted to study. We used four main criteria to ensure that our dataset contains high-quality packages. The summary statistics of these steps are shown in Table~\ref{tab:selection_step}.

The first criterion in our selection process is to select mature and popular packages. To do so, we chose the top 100 \npm packages in our dataset based on the number of stars they received on Github. We chose to use the number of stars since prior work shows that the number of stars can provide a good proxy for the popularity and maturity of software applications and packages~\cite{BORGES2018112,DabbishCSCW2012}.

Second, we eliminated any packages from the dataset that contain at least one release that is labelled as pre-releases or post-releases. We chose packages that do not have pre-releases or post-releases since this is a good indicator that the developers of those packages are somehow familiar with the \sv practices~\cite{DecanTSE2019}. Also, we eliminated those packages to simplify our classifications process since we would have only the three \sv type as labels in our dataset. 

The third step to select the studied \npm packages was to examine packages with a sufficient number of releases. We filtered out from our dataset any package that does not have at least five releases of each type of the \sv, and in total, the package must have at least 50 releases. We excluded packages with a small number of releases since we wanted to use ML techniques to determine the type of \sv. Thus, we wanted to have a sufficient number of labelled releases so that we could build robust ML classifiers.


\begin{figure}[t]
	\centering
	\includegraphics[width=0.5\textwidth]{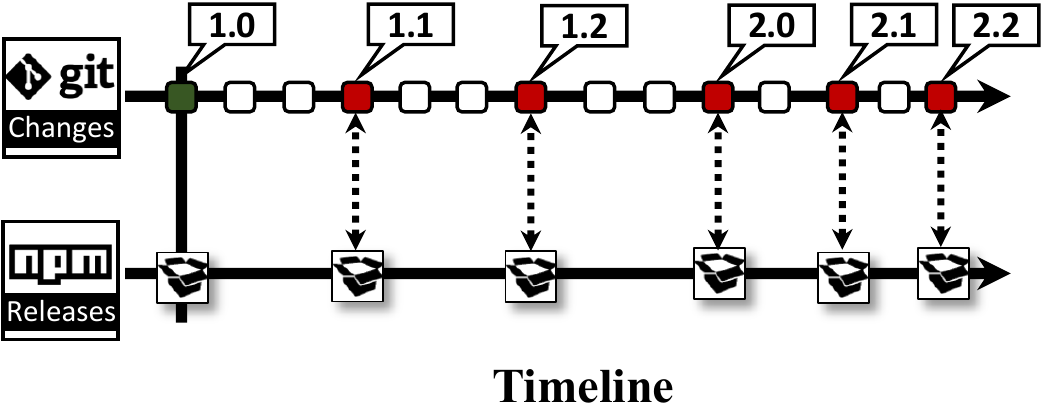}
	\caption{Our approach of identifying the period release history on GitHub.}
	\label{fig:heuristic_approach}
\end{figure}


We finally excluded packages that have any breakage releases identified by developers. It is important to note that we performed this filtration step to ensure that the developers of our studied packages understand \sv and use it adequately in practice. Thus, we had a high-quality labelled dataset. To examine this criterion, for every \npm package in our dataset, we searched on Github for the applications that use these packages. Then, we analyzed the development history of those applications. After that, we examined them to see whether the developers of those applications that use the package had downgraded a version of that package and indicated that they performed the downgrade due to a breakage in the release of the package. Mainly, we analyzed the historical data of these applications and identified the commits where the developers rolled back a version of the selected packages. We then manually examined those commits to determine if developers rolled back a version of the selected packages due to a breaking release that is not correctly specified by the right \sv tag. Finally, we removed any package from our dataset containing at least one case of such a rollback. At the end of this step, we ended up having 36 packages in our dataset.


\subsection{Dataset Preparation}
Once we decided which \npm packages we would use in our study, we cloned them locally and collected their metadata information from the \npm registry. Then, we built a \sv parser to analyze every sequence release of every package to label the release type, whether a release is major, minor, or patch release based on the prior release. For example, suppose a package has a release in an older date that holds the \sv number as \texttt{3.2.6}, and the subsequent release based on the date has the \sv number as \texttt{3.3.6}. In that case, we considered that release as a minor release for that package (i.e., we labelled it as a minor release type). It is worth mentioning that following this process, we were able to identify and eliminate any backport releases from our dataset.

In the next step and since we wanted to extract features based on the source code and the development history of the packages' releases in our study, we needed to have the source code and the development history of each package in our dataset. Therefore, for each package in our dataset, we started by collecting their metadata information and source code from the public repository of \npm. To do so, for each \npm package in our dataset, we downloaded the appropriate `tar' file that contains the source code of every release of that package. In addition, we collected the release date for every release of the packages and the GitHub repository URL of the packages.

Now, we had the source code of each release. Next, we wanted to collect the historical development data from the GitHub repository of each package. We used the provided URL link to the GitHub repository to access the development history. Then, we cloned the GitHub repository of each package and analyzed it. However, we could not clone two package repositories because their GitHub repositories do not exist or are changed to private repositories. In addition, based on our research experience with the \npm registry, we noted that more than one \npm packages could be hosted on the same GitHub repository (i.e., they hosted in monorepo repository). Thus, we manually examined the selected packages and remove three packages from our dataset that their GitHub repository contains more than one \npm packages.

Once we collected the release information from \npm and GitHub repositories, we used a heuristic approach based on the release date to link each release to its development history on the GitHub repository. Figure~\ref{fig:heuristic_approach} shows the overall approach. First, we analyzed the release date from the \npm registry for each package release in our dataset. And then, we extracted all the commits and their metadata. By analyzing the commits, we extracted the commit date. Based on the release date, we identified the first commit and the last commit for each release (i.e., we identified the release timeframe). Now we had the source code and the development history of each package release in our dataset, we analyzed these data to extract a comprehensive set of features. We describe our process for extracting the studied features for \npm packages in our dataset in the next section (Section~\ref{sec:features}).


Table~\ref{tab_studied_packages} presents various statistics of our studied JavaScript packages from \npm. It shows first the name of the package and the number of commits. In addition, the Table shows the total number of releases, the number of analyzed releases of the studied packages, and the percentage of major, minor, and patch releases of the studied packages. In total, there are 31 packages in our dataset.


\subsection{Features for Semantic Versioning Classification}
\label{sec:features}
Since our goal is to perform release-level predictions to determine the \sv type of a new package release, we resorted to using some of the most commonly used release-level features. Some of these features were used in prior software engineering tasks to identify post-release defects~\cite{ShihabESEM2010} or used to determine crashing releases of mobile apps~\cite{Xia_ESEM2016}. Therefore, we believed that some of these features can be used to determine the level of complexity of a new package release, hence, providing useful information as to determine the type of a new release.

To perform our study of determining the \sv type of a new release, we resorted to using release-level features. In total, we extracted 41 features that are categorized into six dimensions. We distinguished between these feature categories since; 1) it allowed us to observe the contribution of different types of features, and 2) these categories let us organize how we created and interpreted features related to determining the \sv type. In general, we extracted these features from analyzing the source code and the development activities of each new package release in our dataset. Table~\ref{tab:features} presents the names and the definition of the extracted features, and the rationale for examining them. In the following subsections, we presented the detailed process of extracting the studied features in each of the six dimensions.

\noindent\textbf{Change Type Features:} Change type features present the source code elements that may impact the \sv type of a new package release. To extract change type features, we resorted to using source code analysis to calculate these features (described in Table~\ref{tab:features}). Thus, we analyzed the changes made after each release and extracted fine-grained source code change types. To extract the features from code changes, we used the GumTree code differencing algorithm~\cite{FalleriGumTreeASE2014}. GumTree takes as input the pair of revision files and creates two Abstract Syntax Trees (ASTs) that are used to compare those different revisions. As a result, GumTree outputs a list of fine-grained source code changes (e.g., an update in a method invocation or rename). Then, we wrote scripts that extract the fine-grained source code change types based on the GumTree algorithm.

To extract change types features based on code that happened in each release, we needed to have the complete version of the JavaScript files before and after the release. To do so, we ran the diff command line between two consecutive releases. Then, we extracted all the JavaScript files where the files' names have a \texttt{.js} extension (i.e., JavaScript source file). Once we had the two revisions of each changed file in two consecutive releases, we ran the GumTree tool on them. After that, we analyzed the results of GumTree to extract the change-type features. Since the GumTree tool's output is in a JSON format, we parsed the resulting JSON files to retrieve the differences between the before and after files versions. Based on this step's results, we counted the number of element changes in every two revisions of files and then summed up them to get a change type value for each release.




\noindent\textbf{Dependencies Features:} Dependency features present the dependencies change activities that occurred while developing a new package release. To calculate the dependency-related features, we analyzed the changes that happened to the {package.json} file. First, we analyzed the package.json file since it is the configuration file used in the studied packages to manage and configure dependencies. Then, we calculated the number of commits that touch the package.json file and the number of commits that added, deleted, updated packages in the package.json file. We built a tool that analyzes the package.json file at every release and compares it with the previous releases to identify dependencies that were changed.

\noindent\textbf{Complexity and Code Features:} Complexity and code features represent the package's source code changes in each release. To calculate the complexity and code features (e.g., the difference average of Cyclomatic and the total line of code added and deleted) for each examined release in our dataset, we analyzed the release's source code and computed the diff of the analyzed release with the previous releases. To achieve this, we ran the Understand tool~\cite{SciTools2Online} on every release for the examined packages in our dataset and calculated the difference between the current release and the one before.

\noindent\textbf{Time Feature:} The time feature presents the time that a new release takes to be developed and published. We counted the number of days a new release takes to be published since the previous release date to calculate the time feature.

\noindent\textbf{Development Features:} Development features present the development activities performed during the development of a new release of a package. To calculate the development features, we analyzed the GitHub repository of each package in our dataset. Then we measured the number of commits, unique developers, open issues, closed pull requests, and open pull requests that occurred during that release development timeframe.

\noindent\textbf{Textual Features:} Text features present extracted information from the commit change logs that the developers have written during the development of a new release. To extract the text features, we analyzed the commit message and looked for specific keywords, ``major'', ``patch'', ``break'', and then counted the number of commits containing these keywords in each release. As for the identify bug-fixing commits, we used a well-known approach that based on examining the appearance of a pre-defined set of keywords that include ``bug'', ``fix'', ``defect'', ``error'', ``issue'', and their variants in commit messages~\cite{sliwerski2005changes,williams2008szz}. Then, we counted those commits in every studied release.

\begin{table*}[h]
	\centering
	\caption{Features used to determine the \sv type of a new \npm package release.}
	\label{tab:features}
	  \scalebox{0.78}{
\begin{tabular}{p{0.2in}|p{0.35in}|p{3.6in}|p{3.5in}}
\hline


\multirow{2}{*}{\textbf{Dim.}} & \multirow{2}{*}{\textbf{Name}} & \multirow{2}{*}{\textbf{Definition}} & \multirow{2}{*}{\textbf{Rational}} \\
&  &  &  \\
\hline \hline

\multirow{20}{*}{\rotatebox{90}{\textbf{Change type}\hspace{0.8cm}}} &\textit{AJF}& The number of JavaScript files added between two releases. & \multirow{6}{*}{\begin{tabular}[c]{@{}l@{}}
		The releases that modify several JavaScript files, functions or/and \\change the code structure in \npm packages tend to be more major \\releases than being minor or patch releases. Furthermore, these are \\change types that can provide good indications of the semantic ve-\\rsioning type of a new \npm package release. In other words, the re-\\leases that include adding new JavaScript functionalities are not \\small releases that are more likely to be major releases. For exam-\\ple, if there are several JavaScript files that are deleted in a new \\package release, then that release is not expected to be a patch or \\a minor release. Another example, If there are several non-JavaSc-\\ript files are changed (i.e., added, deleted, or modified) in a new \\package release, then the release is likely to be a patch or a minor \\release.
\end{tabular}}
 \\ \cline{2-3}
 & \textit{MJF} & The number of JavaScript files modified between two releases. &   \\ \cline{2-3}
 & \textit{DJF} &  The number of JavaScript files deleted between two releases.&  \\ \cline{2-3}
 & \textit{ANJF} &The number of non-JavaScript files added between two releases. &   \\ \cline{2-3}
 & \textit{DNJF} & The number of non-JavaScript files deleted between two releases.&  \\ \cline{2-3}
 & \textit{MNJF} & The number of non-JavaScript files modified between two releases.&  \\ \cline{2-3}
 & \textit{ADM} &  The number of methods that are added between two releases. &  \\ \cline{2-3}
 & \textit{DEM} &  The number of methods that are deleted between two releases.&  \\ \cline{2-3}
 & \textit{MOM} & The number of methods that are moved between two releases.&  \\ \cline{2-3}
 & \textit{MNC} & The number of methods whose names are changed between two releases.&  \\ \cline{2-3}
 & \textit{MPC} & The number of methods whose input parameters are changed between two releases.&  \\ \cline{2-3}
 & \textit{MPD} &  The number of methods whose input parameters are deleted between two releases.&  \\ \cline{2-3}
 & \textit{MLA} &  The number of logics in methods are added between two releases.&  \\ \cline{2-3}
 & \textit{MLM} &  The number of logics in methods are moved between two releases.&  \\ \cline{2-3}
 & \textit{MLD} &  The number of logics in methods are deleted between two releases.&  \\ \cline{2-3}
 & \textit{GVA} &   The number of global variables added in JavaScript files between two releases.&  \\ \cline{2-3}
 & \textit{GVD} & The number of global variables deleted in JavaScript files between two releases.&  \\ \cline{2-3}
 & \textit{ICC} &  The number of total code comments added between two releases.&  \\ \cline{2-3}
 & \textit{DCC} & The number of total code comments deleted between two releases.&  \\ \cline{2-3}
 & \textit{MCC} & The number of total code comments modified between two releases.&  \\ \hline

\multirow{6}{*}{\rotatebox{90}{\textbf{Dependency}\hspace{0.01cm}}} & \textit{TCPJ} & The number of changes to the package.json file. & \multirow{4}{*}{\begin{tabular}[c]{@{}l@{}}The releases that have more updates to the package dependencies\\list are more likely not to be patch releases. For example, adding \\more dependencies into the package dependencies list in the new \\release can indicate that this release is a major release. Another \\example, the changes that delete more dependencies in the new \\release can indicate a major release rather than a minor or a patch \\release.  
\end{tabular}}\\ \cline{2-3}
 & \textit{PA} & The number of used packages added between two releases.~~~~~~~~~~~~~~~~~~~~~~&  \\ \cline{2-3}
 & \textit{PD} & The number of used packages deleted between two releases. ~~~~~~~~~~~~~~~~~~~~~~&  \\ \cline{2-3}
 & \textit{PU} & The number of used packages' versions changed between two releases.&  \\ \hline

\multirow{7}{*}{\rotatebox{90}{\textbf{Complexity}\hspace{0.1cm}}} & \textit{ACYCD} & The difference average of Cyclomatic between two consecutive releases. & \multirow{3}{*}{\begin{tabular}[c]{@{}l@{}}  We expect that the complexity and code features provide strong \\indicators of the \sv type of the new release. If \\the complexity and the package size change a lot in the new release, \\these changes will likely present the type of \sv \\release. For example, a large diff number of lines between two \\releases indicate that the new release introduces more code and is \\more likely not to be a patch or a minor release.    \end{tabular}} \\ \cline{2-3}
 & \textit{CLCJD} & The difference of  lines of code between two consecutive releases. ~~~~~~~~~~~~~~&  \\\cline{2-3}
 & \textit{CYCD} & The difference Cyclomatic between two consecutive releases. ~~~~~~~~~~~~~~~& \\ \cline{2-3}
  & \textit{LA} & The total line of code added between two releases.&  \\ \cline{2-3}
 & \textit{LD} & The total line of code deleted between two releases. &  \\ \hline
 


\rotatebox{90}{\textbf{Time~~}\hspace{0.0cm}} & \textit{RDTD} & The timestamp difference between two consecutive releases. &  A package release development that takes a long time tends to contains several changes, which is not likely to be patch. \\ \hline

\multirow{7}{*}{\rotatebox{90}{\textbf{~~Development}\hspace{0.0cm}}} & \textit{TCM} & The total number of commits between two releases. & \multirow{6}{*}{\begin{tabular}[c]{@{}l@{}} The \sv type of a new package heavily de-\\pends on the number of development activities in that rele-\\ase. For example, many commits or many numbers of clos-\\ed pull requests happened during the releases; this indicat-\\es that ~this release is not a patch ~release but tends to be a \\major or a minor package release. \end{tabular}} \\ \cline{2-3}
   & \textit{TAU} & The total number of authors made changes between two releases. &  \\ \cline{2-3}
 & \textit{POI} & The total number of open issue between two releases. &  \\ \cline{2-3}
 & \textit{PCI} & The total number of closed issue between two releases. &  \\ \cline{2-3}
 & \textit{PCPR} & The total number of closed pull request between two releases. &  \\ \cline{2-3}
 & \textit{POPR} & The total number of open pull request between two releases. &  \\ \hline


\multirow{6}{*}{\rotatebox{90}{\textbf{\small Textual}\hspace{0.2cm}}} & \textit{NBF} & The total number of bug-fixing commits between two releases. & \multirow{6}{*}{\begin{tabular}[c]{@{}l@{}}
The change ~message contains ~the purpose of this commit. \\For example, commits that several messages contain the k-\\eyword major changes or breakage changes in a release de-\\velopment history provide a high indication that this relea-\\se a major release. On the other hand, releases that have co-\\mmits messages containing the word min-\\or tend to be minor or patch releases.
	\end{tabular}} \\ \cline{2-3}

& \textit{KWM} & The total number of commits that have keyword major in commit message in the release.&  \\ \cline{2-3}
& \textit{KWP} & The total number of commits that have keyword patch in commit message in the release.&  \\ \cline{2-3}
& \textit{KWB} & The total number of commits that have keyword break in commit message in the release.&  \\ \cline{2-3}
& \textit{AML} & The average commit message length in commits happened in the release. &  \\ \hline

\hline
\end{tabular}
}

\end{table*}


\subsection{Classification Algorithms}
\label{sec:classifier}
To perform our classification task, we chose four different machine learning algorithms. In particular, we chose to use XGBoost (XGB), Random Forest (RF), Decision Tree (DT), and Logistic Regression (LR) algorithms to classify whether a new package release is a major, minor, or patch. We resorted to using these ML algorithms since they 1) have different assumptions on the examined dataset, 2) show different characteristics in terms of dealing with overfitting and execution speed~\cite{CaruanaICML2006}, and 3) provide an intuitive and straightforward explanation of the classification, which enables developers to easily understand why a decision to determine the type of package release was made~\cite{Kotsiantis2006}. In addition, they have been commonly used in the past in other software engineering studies and datasets~(e., g.~\cite{GhotraICSE2015,kameiTSE2013,BacchelliICSE2012,XiaESEM2016,ThungICSM2012,IbaPPSN1996,HeASEJ2012}). We then compared the performances of these different supervised classifiers to determine the type of release. Now, we briefly described the four examined machine learning algorithms.

{\noindent\textbf{XGBoost (XGB):}} The XGBoost classifier is an extended and innovative application of gradient boosting algorithm proposed by Chen et al.~\cite{TianqiKDD2016}. Gradient boosting is an algorithm in which new models are created that predict the residuals of prior models and then added together to make the final prediction. Models are added recursively until no noticeable improvements can be detected. This approach supports both regression and classification. XGBoost has proven to push the limits of computing power for boosted tree algorithms. Furthermore, prior work showed that applying the XGBoost classifier on software engineering data produced good performance (e.g.,~\cite{esteves2020understanding,Mariano2019ICMLA})


{\noindent\textbf{Random Forest (RF):}} The Random Forest classifier is a type of combination approach, which is bagging and random subsets meta classifier based on a decision tree classifier~\cite{breiman2001random}. Random Forest combines multiple decision trees for prediction. First, each decision tree is built based on the value of an independent set of random vectors. Then, the Random Forest classifier adopts the mode of the class labels output by individual trees. Also, prior work showed that it performs well on software engineering problems~(e.g.,~\cite{RahmanMSR2017,YanTSE2019}).

{\noindent\textbf{Decision Tree (DT):}} 
The decision trees classifier first creates a decision tree based on the feature values of the training data where internal nodes denote the different features~\cite{Quinlan1993}. The branches correspond to the value of a particular feature, and the leaf nodes correspond to the classification of the dependent variable. Then, the decision tree is made recursively by identifying the feature(s) that discriminate the various instances most clearly, i.e., having the highest information gain~\cite{hall2009weka}. Once a decision tree is built, the classification for a new instance is performed by checking the respective features and their values.

{\noindent\textbf{Logistic Regression (LR):}} The Logistic Regression is used to estimate the probability of a binary response based on one or more independent variables (i.e., features). Previous work showed that regression-based classifiers, especially logistic regression, usually achieve high performance on software engineering classification tasks (e.g., \cite{GhotraICSE2015,kameiTSE2013}).




{\textbf{Baseline:}} Finally, to put our ML classification results in perspective, we chose to use a simpler classifier as a baseline. In our study, we decided to use the ZeroR (ZR) classifier, which is a primitive classifier~\cite{WekaManu91online}. It basically predicts the majority class in the training data for all cases in the test data without considering the independent features. 


\subsection{Training and Testing Classifiers}
\label{traning_testing}

To conduct our experiments and answer our research questions, we constructed an ML pipeline to build three different groups of classifiers. We first built within-package classifiers where we used all the six dimensions of features to train and test data from one package. Second, we built within-package classifiers for each package based on each feature's dimensions (i.e., for each package, we built six classifiers). Finally, we built cross-package classifiers, where for each package, a cross-package classifier is trained on data from all packages except one and tested on the remaining one package. 

Since, in our case, we have a multi-classes ML problem (e.g., as a major, minor, patch), we formalized our ML problem to binary classification problems. In another word, we used a one-versus-the-rest approach~\cite{murphy2012machine}. We used one-versus-the-rest classifiers to ease the interpretation of our classifiers' outcomes. In our study, we built three one-versus-the-rest classifiers for each new release type: a major release or not, a minor release or not, and a patch release or not. Thus, this requires creating three different ML classifiers and training each of them with true positives and true negatives (e.g., true minor releases and not minor releases). Furthermore, to train and test our classifiers, we used the 5-fold cross-validation technique. In each 5-fold cross-validation, we divided the dataset into five folds. Then, four folds are used to train the classifier, while the remaining one fold is used to evaluate the performance of the built classifier. This process is repeated five times so that each fold is used exactly once as the testing set. We resorted to using 5-fold cross-validation to reduce the bias due to random training data selection~\cite{bengio2004no}. We finally reported the average performance across these test runs. The reported results are the average of 5-fold cross-validation, such that each sample in the total dataset was included exactly in one test set. We implemented our examined classifiers using scikit-learn~\cite{pedregosa2011scikit}. We also used the default scikit-learn configuration to set the different parameters of the examined classifiers.


Furthermore, and as it is shown in Table~\ref{tab_studied_packages}, our dataset has on average 10.09\%, 29.72\%, and 60.20\% for major, minor, and patch releases, which indicate that our dataset contains imbalances data. Data imbalance occurs when one class occurs much more than the other in a dataset, which leads to the situation that the trained classifiers will learn from the features affecting the majority cases than the minority cases~\cite{SongTSE2019}. To deal with the imbalance problem in our experiments, we applied the synthetic minority oversampling technique (SMOTE). SMOTE is a method for oversampling and can effectively boost a classifier’s performance in an imbalanced case dataset such as our dataset~\cite{chawla2002smote}. We applied the sampling technique to our dataset since it balances the size of the majority class and allows us to report standard performance and better interpret our results. It is essential to highlight that we only applied this sampling technique to the training dataset. We did not re-sample the testing dataset since we want to evaluate our classifier in a real-life scenario, where the data might be imbalanced.

\subsection{Performance Measures}
To evaluate the performance of the used four machine learning classifiers and compare their performance to our baseline, the ZeroR classifier, we calculated the Area Under the Receiver Operating Characteristic curve (ROC-AUC). ROC-AUC is a well-known evaluation measurement that is considered statistically consistent. In the ROC curve, the true positive rate (TPR) is plotted as a function of the false positive rate (FPR) across all thresholds. More importantly, ROC-AUC is a threshold independent measure~\cite{bradley1997use}. A threshold represents the likelihood threshold for deciding an instance that is classified as positive or negative. Usually, the threshold is set as 0.5, and other performance measures for a classifier, such as the F1-score, heavily depend on the threshold's determination. However, some cases may need to change the threshold, such as the class imbalance case. Thus, we used ROC-AUC to avoid the threshold setting problem since ROC-AUC measures the classification performance across all thresholds (i.e., from 0 to 1). Likewise, ROC-AUC has the advantage of being robust towards class distributions~\cite{LessmannTSE2008,NamASE2015}. 

The ROC-AUC has a value between 0 and 1, where one indicates perfect classifications results and zero indicates completely wrong classifications. It is important to note that prior work shows that achieving a 0.5 ROC-AUC value indicates that the classifier performance is as good as random, while the ROC-AUC value equal to or more than 0.7 indicates an acceptable classifier performance using software engineering datasets~\cite{NamASE2015,LessmannTSE2008,YanTSE2019}.


	\section{Case Study Results}
	\label{sec:case_study_results}
\begin{table*}[t]
	\centering
	\caption{The performance of the examined four ML classifiers for determining the release type - major, minor, and patch. The results are reported for XGBoost (XGB), Random Forest (RF), Decision Tree (DT), and Logistic Regression (LR). In addition, the Table shows the results of our baseline classifier, which is the ZeroR (ZR). The best performance values are highlighted in bold.} 
	\label{tab_rq1_results}
	  \scalebox{0.88}{
\begin{tabular}{p{1.25277in}|p{0.23in}|p{0.23in}|p{0.2in}|p{0.23in}|p{0.23in}||p{0.23in}|p{0.23in}|p{0.20in}|p{0.23in}|p{0.23in}||p{0.23in}|p{0.23in}|p{0.2in}|p{0.23in}|p{0.23in}}
\hline


\multirow{4}{*}{\textbf{Packages}} & \multicolumn{5}{c||}{\multirow{2}{*}{\textbf{\underline{~~~~~~Major~~~~~~}}}} & \multicolumn{5}{c||}{\multirow{2}{*}{\textbf{\underline{~~~~~~Minor~~~~~~}}}} & \multicolumn{5}{c}{\multirow{2}{*}{\textbf{\underline{~~~~~~Patch~~~~~~}}}} \\
& \multirow{2}{*}{\textbf{XGB}} & \multirow{2}{*}{\textbf{RF}} & \multirow{2}{*}{\textbf{ZR}} & \multirow{2}{*}{\textbf{DT}} & \multirow{2}{*}{\textbf{LR}} & \multirow{2}{*}{\textbf{XGB}} & \multirow{2}{*}{\textbf{RF}} & \multirow{2}{*}{\textbf{ZR}} & \multirow{2}{*}{\textbf{DT}} & \multirow{2}{*}{\textbf{LR}} & \multirow{2}{*}{\textbf{XGB}} & \multirow{2}{*}{\textbf{RF}} & \multirow{2}{*}{\textbf{ZR}} & \multirow{2}{*}{\textbf{DT}} & \multirow{2}{*}{\textbf{LR}} \\
  &  &  &  &  &  &  &  &  &  &  &  &  &  &  &  \\

 \hline \hline

sweetalert2                                    &0.85&\textbf{0.92}&0.44&0.59&0.76&     \textbf{0.73}&0.71&0.49&0.56&0.59&         \textbf{0.74}&\textbf{0.74}&0.52&0.61&0.65\\
renovate                                          &\textbf{0.93}&0.89&0.43&0.49&0.67&    \textbf{0.87}&0.84&0.50&0.69&0.66&        \textbf{0.86}&0.81&0.51&0.71&0.67\\
speakingurl                                     &\textbf{0.73}&\textbf{0.73}&0.50&0.60&\textbf{0.73}&                    0.44&0.34&0.53&0.46&\textbf{0.65}&        \textbf{0.74}&0.72&0.45&0.64&0.63\\
license-checker                             &0.62&\textbf{0.64}&0.47&0.52&0.46&     \textbf{0.59}&0.50&0.49&0.52&0.39&        0.73&\textbf{0.75}&0.52&0.63&0.62\\
bittorrent-dht                                &0.86&\textbf{0.87}&0.42&0.54&0.65&     0.51&\textbf{0.61}&0.54&0.49&0.59&          0.67&\textbf{0.74}&0.48&0.60&0.53\\
nes                                                   &0.48&0.42&0.44&\textbf{0.56}&{0.49}&     \textbf{0.82}&0.76&0.51&0.66&0.63&          \textbf{0.68}&0.66&0.53&0.60&0.67\\
box-ui-elements                           &0.84&\textbf{0.89}&0.42&0.64&0.68&     \textbf{0.68}&0.60&0.49&0.61&0.61&          0.74&0.76&0.53&0.63&\textbf{0.83}\\
sku                                                   &\textbf{0.86}&0.73&0.50&0.60&0.50&     \textbf{0.79}&0.75&0.51&0.66&0.56&          \textbf{0.78}&0.70&0.44&0.67&0.64\\
mongo-sql                                      &\textbf{0.83}&0.68&0.48&0.70&0.50&     0.64&\textbf{0.78}&0.43&0.61&0.72&          0.65&\textbf{0.68}&0.43&0.62&0.62\\
pacote                                             &\textbf{0.93}&0.90&0.52&0.78&0.84&     \textbf{0.82}&0.81&0.46&0.61&0.77&           0.85&\textbf{0.87}&0.45&0.71&0.66\\
seek-style-guide                          &\textbf{0.72}&0.62&0.48&0.55&0.42&      \textbf{0.76}&\textbf{0.76}&0.51&0.63&0.55&           \textbf{0.75}&0.73&0.49&0.67&0.61\\
nightwatch-cucumber               &0.76&\textbf{0.81}&0.48&0.56&0.46&       0.73&\textbf{0.80}&0.53&0.61&0.65&           0.76&\textbf{0.83}&0.50&0.70&0.65\\
zapier-platform-cli                      &\textbf{0.87}&0.85&0.54&0.75&0.69&      \textbf{0.78}&0.75&0.54&0.57&0.65&          0.82&\textbf{0.83}&0.48&0.73&0.64\\
patchbay                                      &\textbf{0.68}&0.60&0.51&0.45&0.33&        0.69&\textbf{0.72}&0.47&0.62&0.59&          0.68&\textbf{0.73}&0.48&0.60&0.57\\
module-deps                                &0.75&\textbf{0.80}&0.57&0.51&0.64&       \textbf{0.65}&0.60&0.47&0.59&0.43&          \textbf{0.68}&0.61&0.48&0.59&0.49\\
turtle.io                                         &0.77&\textbf{0.88}&0.53&0.56&0.79&        \textbf{0.80}&0.76&0.49&0.58&0.64&          0.81&\textbf{0.85}&0.54&0.72&0.77\\
rtcpeerconnection                     &\textbf{0.75}&0.62&0.50&0.57&0.71&          \textbf{0.59}&0.55&0.54&0.55&0.57&          0.62&0.44&0.51&0.55&\textbf{0.63}\\
react-isomorphic-render         &\textbf{0.82}&0.80&0.55&0.54&0.59&         0.74&\textbf{0.75}&0.48&0.55&0.47&          \textbf{0.80}&\textbf{0.80}&0.51&0.73&0.60\\
rtc-quickconnect                        &0.78&\textbf{0.85}&0.58&0.60&0.78&        \textbf{0.72}&0.66&0.51&0.66&0.58&           \textbf{0.78}&\textbf{0.78}&0.50&0.64&0.63\\
terrestris/react-geo                   &0.64&\textbf{0.75}&0.45&0.50&0.53&        \textbf{0.67}&0.66&0.45&0.61&0.60&           \textbf{0.71}&0.66&0.58&0.63&0.62\\
eslint-config-canonical             &0.82&\textbf{0.83}&0.50&0.75&0.56&        0.64&\textbf{0.69}&0.48&0.55&0.49&          \textbf{0.74}&\textbf{0.74}&0.51&0.63&0.58\\
repofs                                            &0.80&\textbf{0.91}&0.47&0.57&0.57&         0.72&\textbf{0.84}&0.49&0.58&0.42&           0.76&\textbf{0.83}&0.49&0.65&0.58\\
penseur                                        &0.64&\textbf{0.76}&0.49&0.49&0.61&         \textbf{0.68}&0.66&0.57&0.58&0.56&           \textbf{0.75}&\textbf{0.75}&0.45&0.71&0.73\\
octokit/routes                            &\textbf{0.82}&0.65&0.49&0.68&0.65&         \textbf{0.71}&0.59&0.52&0.55&0.56&            0.63&\textbf{0.67}&0.53&0.57&0.57\\
socketcluster-server                 &0.78&\textbf{0.80}&0.42&0.58&0.73&         0.45&0.45&0.46&\textbf{0.49}&0.46&          0.70&\textbf{0.73}&0.47&0.68&0.63\\
oui                                                  &0.88&\textbf{0.96}&0.54&0.69&0.65&        \textbf{0.95}&0.84&0.44&0.70&0.64&           0.91&\textbf{0.94}&0.55&0.83&0.75\\
express-processimage              &\textbf{0.67}&0.39&0.46&0.48&0.47&        \textbf{0.62}&0.61&0.46&0.60&0.51&             \textbf{0.69}&0.68&0.50&0.59&0.61\\
octokit/fixtures                            &{0.75}&0.71&0.57&\textbf{0.77}&0.61&          \textbf{0.74}&0.70&0.52&0.70&0.65&             \textbf{0.70}&0.62&0.48&0.61&0.52\\
jsonrpc-bidirectional                 &\textbf{0.62}&0.50&0.49&0.61&0.53&        0.63&0.59&0.58&0.58&\textbf{0.67}&             0.57&\textbf{0.62}&0.48&0.51&0.60\\
reactive-di                                   &\textbf{0.84}&0.80&0.43&0.66&0.69&        0.56&\textbf{0.59}&0.52&0.46&0.44&            \textbf{0.75}&0.73&0.49&0.63&0.70\\
rtc-signaller                                &0.81&\textbf{0.85}&0.59&0.51&0.59&          0.63&\textbf{0.64}&0.57&0.61&0.57&             \textbf{0.80}&0.76&0.52&0.64&0.65\\ \hline
\textbf{Average}                      &\textbf{0.77}&0.76&0.49&0.59&0.61&\textbf{0.69}&0.67&0.50&0.59&0.58&\textbf{0.74}&0.73&0.50&0.65&0.63\\
\textbf{Median}                       &0.78&0.80&0.49&0.57&0.61&0.69&0.69&0.50&0.59&0.59&0.74&0.74&0.50&0.63&0.63\\
\hline

\textbf{Relative ROC-AUC}                       &1.58$X$&1.55$X$&--&1.21$X$&1.25$X$&1.38$X$&1.36$X$&--&$1.18X$&1.15$X$&1.49$X$&1.48$X$&--&1.31$X$&1.28$X$\\
\hline
	\end{tabular}
}
\end{table*}

In this section, we presented our case study results for our three research questions. For each research question, we presented the motivation for the question, the approach to answering the question, and the results.

\subsection{\rqi}
\noindent\textbf{Motivation:}
Prior work showed that determining the type of new package release is challenging~\cite{BogartFSE2016}. Even though prior work proposed techniques to detect semantic breaking API changes through static analysis for languages such as Java~\cite{XavierSANER2017,raemaekers2017semantic}, such techniques require a clear definition of the public and private API. Such a distinction does not explicitly exist in many dynamic languages such as JavaScript. In this question, we wanted to effectively determine the \sv type of a new JavaScript package release. Therefore, automatically determining the type of semantic versioning can help guide package maintainers on deciding the versioning type on a new release. In this RQ, we aimed to examine the use of machine learning techniques.


\noindent\textbf{Method:} For each package in our dataset, we used the extracted {41} release-level features that are presented in Table~\ref{tab:features} to train the four classifiers to determine whether a new package release is a major, minor, or patch release. We had reformulated this classification task into a one-versus-the-rest classification problem since this is a multi-class classification problem~\cite{murphy2012machine}. We used one-versus-the-rest classifiers since it would help us adequately interpret our classifiers' results. We had a one-versus-the-rest classifier for each new release type: a major release or not, a minor release or not, and a patch release. Thus, we built three different classifiers for each release type where the true positives will be the examine release type (e.g., true minor releases and not minor releases).


After that, for each package, we used 5-fold cross validation~\cite{bengio2004no}. First, we divided the dataset for each package into {five} folds. Then, we used {four} folds (i.e., 80\% of the data) to train the four ML classifiers and used the remaining one fold (i.e., 20\% of the data) to evaluate the performance of the classifiers. We ran this process five times for each fold (i.e., 1x5-folds). In our study, we used the four ML classifiers described in Section~\ref{sec:classifier} that are XGBoost, Random Forest, Decision Tree, and Logistic Regression.

Finally, to evaluate and compare the performance of the four ML classifiers in determining the \sv type of a new package release, we computed the Area Under the Receiver Operating Characteristic curve (ROC-AUC). Then, to come up with one value for the five runs, we calculated the average of the evaluation measurement for 5-folds five times (i.e., 1x5-fold) for every package in our examined dataset.

\begin{table}[t]
	\centering
	\caption{Mann-Whitney Test (\textit{p}-value) and Cliff's Delta (\textit{d}) for the results of the four classifiers vs. the baseline classifiers for the tree different \sv release types.}
	\label{tab:statistic_differentces_rq1}
	\begin{tabular}{l|r|l|r|l|r|l}
		\hline
		\multirow{2}{*}{\textbf{ML}} & \multicolumn{2}{c|}{\textbf{~~~~~\underline{~~Major~~}}} & \multicolumn{2}{c|}{\textbf{~~~~~~\underline{~~Minor~~}}} & \multicolumn{2}{c}{\textbf{~~~~~~~\underline{~~Patch~~}}} \\ 
		& \textbf{\textit{p}-value} & \multicolumn{1}{c|}{\textit{\textbf{d}}} & \textbf{\textit{p}-value} & \multicolumn{1}{c|}{\textit{\textbf{d}}}  & \textbf{\textit{p}-value} & \multicolumn{1}{c}{\textit{\textbf{d}}} \\ \hline \hline
		XGB &7.973e-11  &0.96   &  1.061e-08& 0.85  &  1.468e-11&0.99   \\ 
		RF &  9.392e-09& 0.85  & 1.474e-08 &0.84   & 2.16e-10 &  0.94 \\ 
		DT &3.077e-06  &  0.69  & 3.382e-07 &  0.75 &  4.802e-11&  0.97 \\ 
		LR &4.105e-05  & 0.61 & 0.000254 &  0.54 &2.81e-10  &  0.93 \\ \hline
	\end{tabular}
\end{table}

Since one of the main goals of using machine learning techniques is to help determine the \sv type of new release, we measured how much better the performance of the four used classifiers is compared to the baseline for each package. In our case, the baseline classifier is a classifier that always reports the class of interest based on the majority, which is the ZeroR classifier. In this case, the ZeroR classifier will achieve 100\% recall and precision equal to the rate of examining release type (i.e., major, minor, patch). We followed the previously described process steps to train and test the ZeroR classifier.


Then, we compared the values of ROC-AUC for the four classifiers against the baseline by calculating the relative ROC-AUC (i. e., $Relative~ROC$$-$$AUC$ $=\frac{Examined~Classifier~ROC-AUC}{Baseline~ROC-AUC}$). Relative ROC-AUC shows how much better our classifiers perform compared to the baseline. For instance, if a baseline achieves a ROC-AUC of 10\%, while the XGBoost classifier, for example, achieves a ROC-AUC of 20\%, then the relative ROC-AUC is $\frac{20}{10}=2X$. In other words, the XGBoost classifier performs twice as accurately as the baseline classifier. It is important to note that the higher the relative ROC-AUC value, the better the classifier is in determining the \sv type.

Finally, to examine whether the achieved improvement over the baseline classifier is statistically significant, we performed a non-parametric Mann-Whitney test~\cite{mann1947test} to compare the two distributions for each classifier results in our dataset and determine if the difference is statistically significant, with a $p$-value $<$ 0.05~\cite{mann1947test}. We also used Cliff’s Delta ($d$), a non-parametric effect size measure to interpret the effect size between the four classifier results and our baseline. We then interpreted the effect size value to be small for d $<$ 0.33 (for positive or negative values), medium for 0.33 $\leq d$ $<$ 0.474 and large for $d \geq$ 0.474~\cite{grissom2005effect}.


\noindent\textbf{Result:}
Table~\ref{tab_rq1_results} presents the ROC-AUC values of the four ML classifiers for determining the release type of major, minor, and patch releases. Table~\ref{tab_rq1_results} shows the results for XGBoost (XGB), Random Forest (RF), ZeroR (ZR), Decision Tree (DT), and Logistic Regression (LR) for the 31 studied \npm packages in our dataset. Overall, we observe that for all three different types of the \sv (i.e., major, minor, and patch), the examined four classifiers achieve acceptable performance in terms of ROC-AUC values~\cite{NamASE2015,LessmannTSE2008}.


First, to determine the major release type, Table~\ref{tab_rq1_results} shows that XGBoost classifier achieves ROC-AUC values range between 0.48 and 0.93 with an average ROC-AUC value equal to 0.77 (median$=$0.78). Also, the Random Forest achieves a comparable performance in classifying major release types. The Table shows that Random Forest has an average value of ROC-AUC equal to 0.76. Second, as for the minor releases, we observed that again the XGBoost and Random Forest classifiers perform better than the Decision Tree and Logistic Regression classifiers. Table~\ref{tab_rq1_results} shows that XGBoost and Random Forest have average ROC-AUC values equal 0.69 and 0.67. Lastly, the highest ROC-AUC values for determining the patch release types obtained by the XGBoost classifier range between 0.57 and 0.91, with an average of 0.74 (median$=$0.74). In contrast, the second highest average ROC-AUC for determining the patch release type is achieved by  Random Forest with ROC-AUC values ranging between 0.44 and 0.94 and with an average value of 0.73 (median $=$ 0.74). In general, the achieved ROC-AUC values indicate that the XGBoost classifier effectively determines the different \sv types compared to the other examined ML classifiers.

Furthermore, Table~\ref{tab_rq1_results} shows the average relative ROC-AUC values when comparing the performance of the four classifiers to our baseline. Overall, the computed relative ROC-AUC shows a significant improvement over the baseline. In particular, for all the 31 packages, the XGBoost outperforms the baseline with average relative ROC-AUC values of 1.58$X$, 1.38$X$, and 1.49$X$ for major, minor, and patch release types, respectively.

Finally, Table~\ref{tab:statistic_differentces_rq1} presents the adjusted $p$-values and effect sizes according to the Cliff’s delta ($d$) test. We observed that the differences are statistically significant in the three \sv types and with a large effect size ($d >$ 0.474).


\conclusion{Our machine learning classifiers achieved a promising performance for determining \sv type of a new package release. They also outperformed our baseline classifier in terms of ROC-AUC values. Out of the four examined ML classifiers, XGBoost tended to achieve the best performance with an average ROC-AUC of 0.77, 0.69, and 0.74 for the major, minor, and patch releases. These results translated to an improvement of {58\%, 38\%, and 49\%} compared to our baseline.}


\subsection{\rqii}
\noindent\textbf{Motivation:}
After determining the type of package release with adequate ROC-AUC values and achieving a good improvement compared to our baseline, we are now interested in understanding what dimensions of features impact determining the type of new package releases the most. In our study, we have {41} release-level features grouped into six dimensions. Therefore, being aware of what dimension of features impacts a new release the most can help gain a deeper understanding of these six dimensions. Also, we aim to provide developers with actionable recommendations (i.e., determine the type of new package release). More importantly, in our case, developers can know what dimensions of features they should carefully examine when specifying the new release type.

\begin{figure*}[t]
	\centering
	\begin{subfigure}[b]{0.325\textwidth}
		{\includegraphics[width=\textwidth]{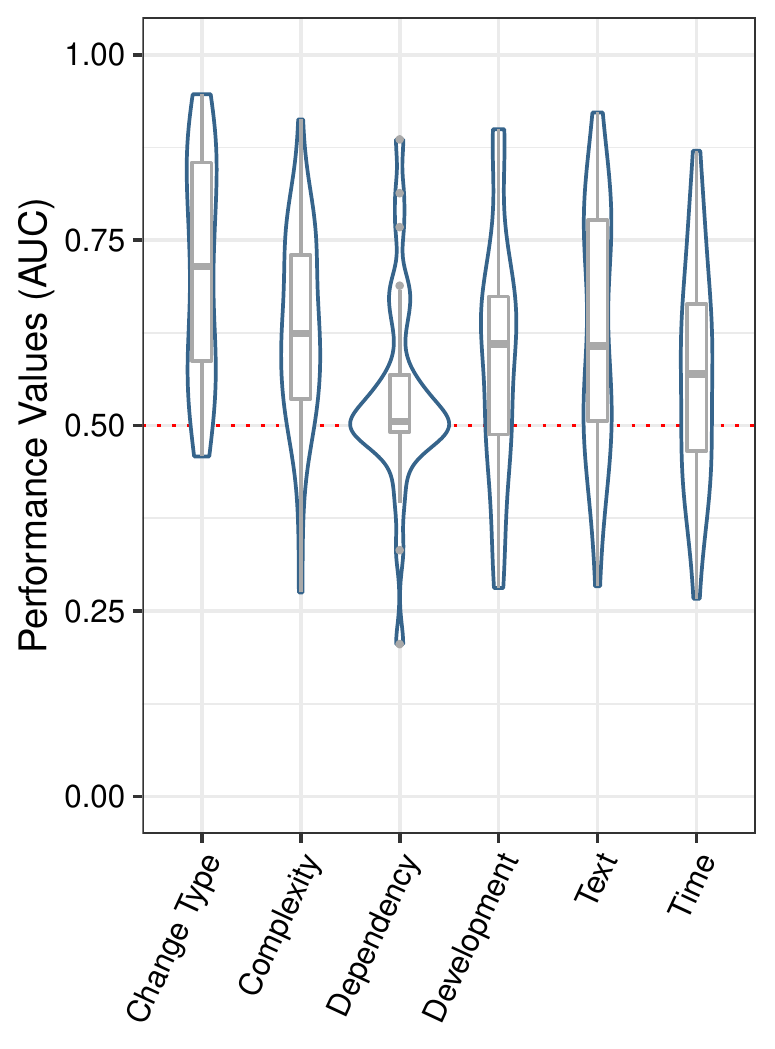}}
		\caption{Major Releases}
		\label{fig:release_npm_all}
	\end{subfigure}
	\begin{subfigure}[b]{0.325\textwidth}
		{\includegraphics[width=\textwidth]{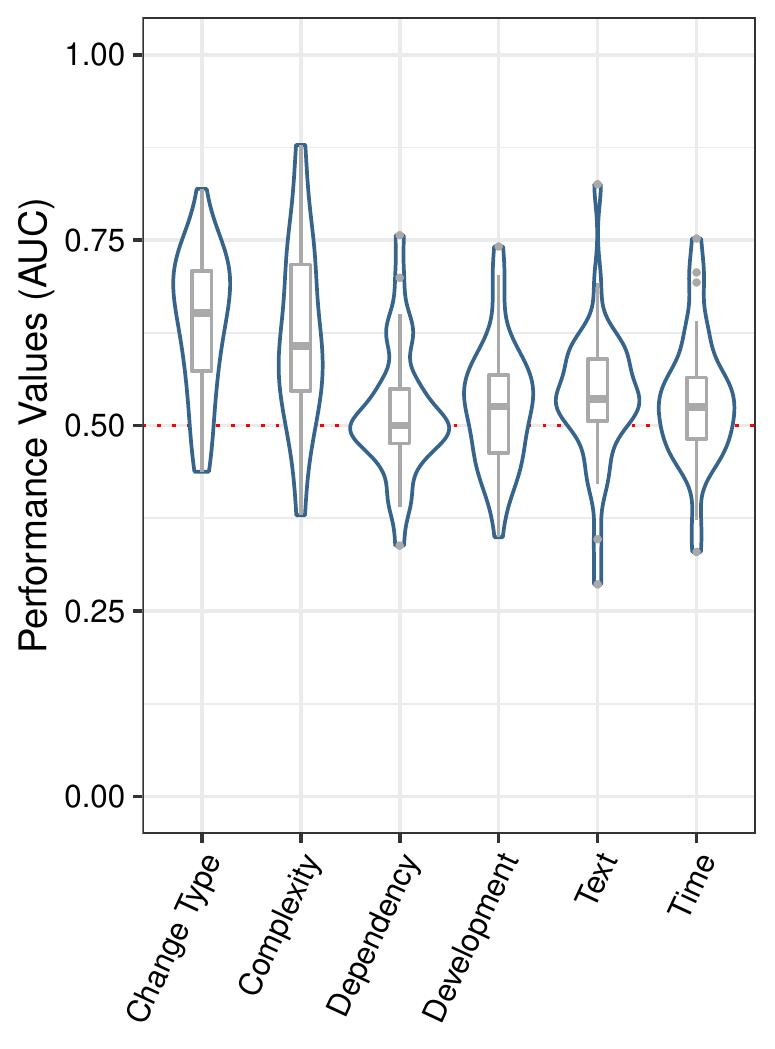}}
		\caption{Minor Releases}
		\label{fig:release_npm_major}
	\end{subfigure}
	\begin{subfigure}[b]{0.325\textwidth}
		{\includegraphics[width=\textwidth]{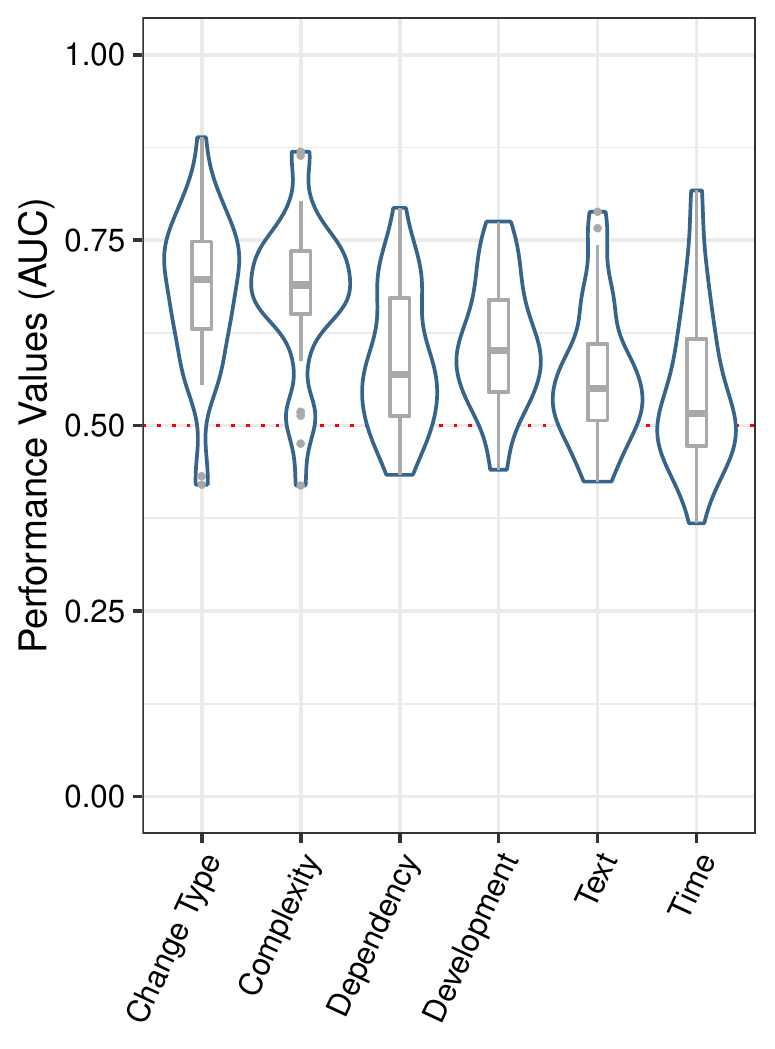}}
		\caption{Patch Releases}
		\label{fig:release_npm_minor}
	\end{subfigure}
	\caption{The distributions of the ROC-AUC values for the different built classifiers. }
	\label{fig:releases_npm}
\end{figure*}

\noindent\textbf{Method:}
To identify the dimension of release-level features that are the most important indicators of determining the \sv type of a new package release, we built several classifiers for each dimension. In particular, for each package release type (i.e., major, minor, patch release), we built six classifiers (one for each dimension of features). In total, we built eighteen classifiers. For example, we built a classifier to determine the major release using the change type dimension of features. To build and evaluate these classifiers, we follow the same steps described in Section~\ref{traning_testing}. Since we found that the XGBoost classifier achieves the best performance in our previous question, we used it as the classifier in this analysis.

Furthermore, to compare and evaluate the performance of the built classifiers based on the different dimensions of features, we again used the well-known evaluation measurement, the ROC-AUC. We then used violin plots to compare the distributions of our results. The vertical curves of violin plots summarize and compare the distributions of different ROC-AUC results.


\noindent\textbf{Result:}
Figure~\ref{fig:releases_npm} shows violin plots of the ROC-AUC values for the built XGBoost classifier for each dimension of features for the three \sv release types. Violin plots are an effective way of presenting the distribution of data. We also superimposed box plots to highlight the key statistics of our results.

From Figure~\ref{fig:releases_npm}, we observed that all the six dimensions of features in our study appear to be important in determining the \sv type of a new package release. However, one dimension of features tended to be a strong indicator of the \sv type of a release, which is the change type dimension. Notably, for the major release type, Figure~\ref{fig:release_npm_all} shows that the best dimension of features to determine the major release type is the change type dimension with an average ROC-AUC value equal to 0.72 (median $=$ 0.72).

As for the minor release, the violin plots in Figure~\ref{fig:release_npm_major} show that the built XGBoost classifiers using the change type dimension outperformed other built classifiers in most of the studied \npm packages. Furthermore, our results showed that the built classifiers based on the complexity and code dimension of features achieved comparable performance to the change type classifiers with average ROC-AUC values equal to 0.70 and 0.68 for classifiers that were built using the change type and complexity and code dimension of features.

For determining the patch release type, from Figure~\ref{fig:release_npm_minor}, we observed that two built classifiers seemed to have comparable results, which are the classifiers that were built using change type and complexity dimensions. These two built classifiers achieved an average ROC-AUC value equal to 0.73 for each. Overall, our built classifiers based on the six dimensions of features in determining the patch release type tended to achieve better performance in terms of average ROC-AUC compared to classifiers built to determine the major and minor release. 

Interestingly, there is some dimension of features that appeared to be a good determine of release type. For example, the dependencies related features appeared to identify patch releases with a good performance. However, classifiers that were built using the dependency dimension of features to determine major and minor releases did not perform as well.


\conclusion{Our investigation showed that the built XGBoost classifiers using the change type dimension of features tended to perform the best when used to determine the \sv release type compared to other built classifiers. However, using all the six dimensions of features still achieved better performance.}

\subsection{\rqiii}
\noindent\textbf{Motivation:}
Building an ML classifier to determine the \sv release type on package-level requires having a sufficient amount of labelled data to train on. However, many packages do not have enough historical labelled data to build a classifier (e.g., newly adopting \sv and/or new packages). Therefore, it would be impossible to train a machine learning classifier to determine \sv type of a new release on data from such packages. In this research question, we investigated to know to what extent and with what performance a \sv type of a new package release can be automatically determined using a cross-package machine learning classification. In addition, answering this question allowed us to evaluate the generalizability of the built classifiers and their applications when applied to other packages.


\begin{table}[t]
	\centering
	\caption{Performance of Cross-packages classification. The results are reported for XGBoost (XGB) and ZeroR (ZR) classifiers.}
		\label{tab_rq3_results}
  \scalebox{0.9}{
	\begin{tabular}{p{1.4in}|r|r||r|r||r|r}
		\hline
%
		
\multirow{2}{*}{\textbf{Package}} & \multicolumn{2}{c|}{\textbf{\underline{~~~Major~~~}}} & \multicolumn{2}{c|}{\textbf{\underline{~~~Minor~~~}}} & \multicolumn{2}{c}{\textbf{\underline{~~~Patch~~~}}} \\		
	
	&\textbf{~~XGB}&\textbf{~~ZR}&\textbf{~~XGB}&\textbf{~~ZR}&\textbf{~~XGB}&\textbf{~~ZR}\\
		
\hline \hline

		sweetalert2&\textbf{0.83}&0.59&\textbf{0.70}&0.48&\textbf{0.75}&0.49\\
		renovate&\textbf{0.58}&0.47&\textbf{0.79}&0.45&\textbf{0.83}&0.51\\
		speakingurl&\textbf{0.71}&0.61&0.56&\textbf{0.62}&\textbf{0.68}&0.39\\
		license-checker&\textbf{0.61}&0.52&\textbf{0.56}&0.33&\textbf{0.72}&0.48\\
		bittorrent-dht&\textbf{0.89}&0.49&0.63&\textbf{0.64}&\textbf{0.75}&0.42\\
		nes&\textbf{0.59}&0.49&\textbf{0.75}&0.49&\textbf{0.75}&0.56\\
		box-ui-elements&\textbf{0.65}&0.57&\textbf{0.62}&0.46&\textbf{0.76}&0.40\\
		sku&\textbf{0.70}&0.51&\textbf{0.80}&0.49&\textbf{0.80}&0.49\\
		mongo-sql&\textbf{0.76}&0.40&\textbf{0.55}&0.54&\textbf{0.60}&0.59\\
		pacote&\textbf{0.92}&0.47&\textbf{0.86}&0.54&\textbf{0.90}&0.52\\
		seek-style-guide&\textbf{0.64}&0.48&\textbf{0.75}&0.46&\textbf{0.77}&0.48\\
		nightwatch-cucumber&\textbf{0.78}&0.53&\textbf{0.80}&0.58&\textbf{0.82}&0.53\\
		zapier-platform-cli&\textbf{0.82}&0.43&\textbf{0.75}&0.53&\textbf{0.82}&0.42\\
		patchbay&\textbf{0.53}&0.51&\textbf{0.77}&0.53&\textbf{0.76}&0.56\\
		module-deps&\textbf{0.82}&0.62&\textbf{0.53}&0.50&\textbf{0.61}&0.49\\
		turtle.io&\textbf{0.88}&0.46&\textbf{0.82}&0.52&\textbf{0.88}&0.44\\
		rtcpeerconnection&\textbf{0.86}&0.59&\textbf{0.56}&0.45&\textbf{0.63}&0.49\\
		react-isomorphic-render&\textbf{0.66}&0.62&\textbf{0.59}&0.57&\textbf{0.63}&0.44\\
		rtc-quickconnect&\textbf{0.84}&0.45&\textbf{0.62}&0.36&\textbf{0.70}&0.58\\
		terrestris/react-geo&\textbf{0.76}&0.53&\textbf{0.65}&0.63&\textbf{0.74}&0.59\\
		eslint-config-canonical&\textbf{0.70}&0.56&\textbf{0.68}&0.41&\textbf{0.78}&0.42\\
		repofs&\textbf{0.86}&0.62&\textbf{0.78}&0.41&\textbf{0.84}&0.49\\
		penseur&\textbf{0.82}&0.28&\textbf{0.57}&0.46&\textbf{0.72}&0.50\\
		octokit/routes&\textbf{0.61}&0.44&\textbf{0.70}&0.64&\textbf{0.63}&0.55\\
		socketcluster-server&\textbf{0.70}&0.52&\textbf{0.61}&0.57&\textbf{0.75}&0.50\\
		oui&\textbf{0.79}&0.63&\textbf{0.58}&0.52&\textbf{0.71}&0.50\\
		express-processimage&\textbf{0.69}&0.45&\textbf{0.69}&0.56&\textbf{0.72}&0.53\\
		octokit/fixtures&\textbf{0.78}&0.52&\textbf{0.86}&0.55&\textbf{0.82}&0.46\\
		jsonrpc-bidirectional&\textbf{0.62}&0.61&\textbf{0.70}&0.54&\textbf{0.73}&0.45\\
		reactive-di&\textbf{0.80}&0.47&\textbf{0.60}&0.49&\textbf{0.74}&0.48\\
		rtc-signaller&\textbf{0.84}&0.50&\textbf{0.75}&0.55&\textbf{0.79}&0.47\\
		\hline
		
		\textbf{Average}&\textbf{0.74}&0.52&\textbf{0.68}&0.51&\textbf{0.75}&0.49\\
		\textbf{Median}&0.76&0.51&0.69&0.52&0.75&0.49\\ \hline
		
		\textbf{Relative ROC-AUC}&1.5$X$&-&1.4$X$&-&1.5$X$&-\\
		\hline

	\end{tabular}
}
%

%
\end{table}

\noindent\textbf{Method:} To better understand the generalizability of the performance achieved by the training classifier on data from one package and apply it to another package, we conducted a cross-packages validation. In particular, we experimented with $n$ fold cross-packages validation, where $n$ is the number of packages in our dataset (i.e., in our dataset, we have 31 packages). We conducted an experiment that trains a classifier on data from thirty packages and uses the built classifier to determine the type of \sv in the remaining one package, similar to the method used in prior work~\cite{Bacchelli2012ICSE,Takafumi2014MSR,Abdalkareem2020TSE}. We repeated this process 31 times, one for each package in our dataset. To build the classifier, we trained the XGBoost machine learning classifiers following the same approach described earlier in Section~\ref{traning_testing}. Once again, we employed the well-known evaluation measurement where we computed ROC-AUC values to measure the performance of the generated classifiers. Finally, to examine the cross-packages classifier's performance with respect to our baseline, which is the ZeroR classifier, we computed the relative ROC-AUC values.

\noindent\textbf{Result:}
Table~\ref{tab_rq3_results} presents the results of our experiment. It shows the ROC-AUC values for each package for the different \sv types. In general, we observed that the built cross-packages classifiers achieved good performance. The built classifiers have average ROC-AUC values of 0.74, 0.68, and 0.75 for the major, minor, and patch releases. With an average ROC-AUC score equal to 0.74 (median$=$0.75), the cross-packages classifier performs significantly high when it is used to determine the major release type. For example, seventeen packages in our dataset have ROC-AUC values greater than 0.75, which is an acceptable performance~\cite{NamASE2015,LessmannTSE2008,YanTSE2019}. We also observed similar performance for determining minor and patch release types.

Moreover, we compared the performance of the cross-packages classifiers to the baseline for all the three \sv release types (i.e., major, minor, and patch). Our results showed that cross-packages classifiers show an improvement of 50\%, 40\%, and 50\% on average over the baseline for the major, minor, and patch \sv release type.

\begin{table}[t]
	\centering
	\caption{Mann-Whitney Test (\textit{p}-value) and Cliff's Delta (\textit{d}) for the results of XGBoost vs. ZeroR classifiers for the tree different version types.}
	\label{tab:compare_general_classifier}
	\begin{tabular}{l|l|l}
		\hline
		\textbf{Version type} & 
		\multicolumn{1}{c|}{\textbf{\textit{p}-value}} & \multicolumn{1}{c}{\textbf{\textit{d}}} \\ \hline \hline
		Major & 4.982e-10  & 0.92 (large)  \\ 
		Minor & 1.42e-08& 0.84 (large) \\ 
		Patch & 1.353e-11 &  1.00 (large)\\ \hline
	\end{tabular}
\end{table}

Finally, we investigated whether the achieved improvements by the built classifiers over the baseline classifiers for the different \sv types are statistically significant. Table~\ref{tab:compare_general_classifier} shows the \textit{p}-values and effect size values. It shows that for all \sv types, the differences are statistically significant, having \textit{p}-values $<$ 0.05. Also, the effect size values are large. These results showed that cross-packages outperform the performance of the cross-package baseline classifier with statistically significant results.

\conclusion{Our results indicated that cross-package machine learning classifiers can provide comparable performances to within-package classifiers for determining the \sv type. For all packages in our dataset, cross-package classifiers achieved average ROC-AUC values of 0.74, 0.68, and 0.75 with an overall improvement over the baseline classifiers with relative ROC-AUC equal to 50\%, 40\%, and 50\% for major, minor, and patch release.}

	
	\section{Related Work}
	\label{sec:related_work}
	In this paper, we proposed using machine learning techniques to effectively determine the \sv type of \npm packages. Thus, our work is mainly related to two areas of prior studies; work related to the use of \sv and work related to identifying breakage changes in third-party packages.

\textbf{Semantic versioning:}
Due to the importance of \sv, several studies have examined it. One of the first works that looked at the use of \sv is the work by~\citet{raemaekers2017semantic}. They investigated the use of \sv in the dataset of 22K Java packages published on Maven that span for seven years. Their results showed that breaking changes occur in 30\% of the studied releases, including minor releases and patches. Thus, several packages used strict dependency constraints, and package maintainers avoid upgrading their dependencies. In addition,~\citet{Kula_EMSE2017} found that developers tend not to update their depend on packages even though these updates are related to the addition of new features and patches to fix vulnerabilities. Interestingly,~\citet{raemaekers2017semantic}'s approach relies on a tool called tclirr, which detects breaking API changes through static analysis of Java code. While a similar tool could be developed for other languages, it requires a clear separation between the public and private API. Such a distinction does not explicitly exist in dynamic languages such as JavaScript, making the accurate detection of breaking changes much more difficult. Moreover, fundamental differences, such as dynamic versus static typing or the language's dynamic nature, between JavaScript and other programming language such as Java make the studies on this language difficult.

\citet{DietrichMSR2019} also studied large dependencies in seventeen package manager ecosystems found that many ecosystems support flexible versioning practices and that the adoption of \sv is increasing. In the same line,~\citet{DecanTSE2019} empirically studied \sv compliances in four ecosystems (Cargo, npm, Packagist, and Rubygems) by analyzing the packages dependency constraints. Their findings showed that the proportion of compliant dependency constraints increases over time in all studied ecosystems.

In the same direction,~\citet{WitternMSR2016} studied the evolution of a subset of JavaScript packages in \npm, analyzing characteristics such as their dependencies, update frequency, and \sv number. They observed that the versioning conventions that maintainers use for their packages are not always compatible with semantic versioning. Also,~\citet{BogartFSE2016} conducted a qualitative comparison of \npm, CRAN, and Eclipse, to understand the impact of community values, tools, and policies on breaking changes. They found two main types of mitigation strategies to reduce the exposure to changes in dependencies: limiting the number of dependencies and depending only on ``trusted packages''. In a follow up work, they interviewed more than 2,000 developers about values and practices in 18 ecosystems~\cite{ChristopherOnlineSurvey}. Among other findings, they observed that package maintainers are frequently exposed to breaking changes and mainly discover them at build time. 


Our work is motivated by these prior aforementioned research efforts. The difference is that our work focuses on proposing a machine learning classifiers to identify the \sv type of a new \npm package release.

\textbf{Identifying breakage changes in third-party packages:}
Several studies investigated API evolution and stability and proposed techniques to detect breakage changes~\cite{Mostafa2017,Xavier_SANER2017,Dig2006AES,Kapur2010RR,AbbasTSE2021}.

\citet{mujahidusingMSR2020} proposed the idea of using other's tests to identify breaking changes of JavaScript packages. They examined the accuracy of their proposed approach on ten cases of breaking updates. Their experimental results showed that their approach identified six breaking updates. Similarly,~\citet{Xavier_SANER2017} performed a large-scale analysis on Java packages. Their results showed that 14.78\% of the API changes are incompatible with previous versions. They also found that packages with a higher frequency of breaking changes are larger, more popular, and more active. Also, ~\citet{Businge_ICSM2012,Businge_SQJ2015} studied Eclipse interface usage by Eclipse third-party plug-ins and evaluated the effect of API changes and non-API changes. ~\citet{MostafaISSTA2017} detected backward compatibility problems in Java packages by performing regression tests on version pairs and by inspecting bug reports related to version upgrades. The similarity between our work and these aforementioned work is the idea of identifying the type of changes in a new package release. However, to the best of our knowledge, our work is the first work to investigated the use of ML technique.

	\section{Threats to Validity}
	\label{sec:threats_to_validity}
	There are few important limitations to our work that need to be considered when interpreting our findings. In this section, we described the threats to the validity of our study.

\textbf{Internal validity:}
Threats to internal validity concerns with factors that could have influenced our study setup. 
First, we used the extracted AST difference between two source codes to extract the change type features. 
To do this, we used GumTree differencing algorithm~\cite{FalleriGumTreeASE2014}. Thus, we might be limited by the accuracy and correctness of this tool. However, previous studies used GumTree for calculating differences between two source codes for different studies. It is also mentioned in the documentation of GumTree that the algorithm is prone to some errors in the context of JavaScript, so it might miss some instances when extracting the difference of JavaScript source codes. For parsing the result of GumTree tool, we developed a parser to extract fine-grained source code changes. This process could result in some errors. Thus, we manually analyzed randomly selected 300 change types to mitigate this threat, and our manual examination shows that the implemented parser correctly extracts all the cases.

In addition, to answer our research questions and to extract the complexity and code dimension of features between two consecutive releases, we used the Understand tool~\cite{SciTools70online}. Therefore, we were limited by the accuracy of the Understand tool. That said, the Understand tool is a widely used analysis tool in both research and industry~\cite{AbdalkareemFSE2017,rahman2019modular,castelluccio2019empirical,ahasanuzzaman2020studying}. Also, a recent study showed that the Understand tool analyzes JavaScript code with good accuracy~\cite{ChowdhuryTSE2021}, which mitigate such a threat.

\textbf{Construct validity:} Threats to construct validity considers the relationship between theory and observation, in case the measured variables do not measure the actual factors.
The labeled package releases (i.e., patch, minor, or major) that we examined are releases that are explicitly marked as so by the package developers in our dataset. In some cases, developers might mislabel the releases. To mitigate this threat, we have applied different filtration criteria (see Section~\ref{TestDatset}) that include selecting mature and popular packages. Also, we filtered out any package that their users reported it to has at least one breakage release but their developers tagged it a minor or patch release. 

	
	


Also, to extract the development features, we opted for analyzing the commits in the Git system. Similar to prior work (e.g.,~\cite{KhomhEMSE2015,Souza}) to identify those commits between two consecutive releases, we consider all commits occurred in the main trunk of the versioning system based on the release date. It is worth mentioning that these dates could be approximations, as developers could start working on the release even before it is issued.

\textbf{External validity:}
Threats to external validity concern the generalization of our findings. 
Our dataset only consists of JavaScript packages, which are published on the \npm  package manager. Hence, our findings might not hold for packages published on other package managers and written in different programming languages. That said, prior work~(e.g., \cite{DecanEMSE2019}) showed that \npm packages are commonly used, and \npm is one of the largest and rapidly growing package managers, which make it the ideal case to study.

In this study, we performed a combination of feature extraction both from code changes and development history from JavaScript open-source packages, and the method used to extract the studied features is specific to JavaScript, so our classifiers might not be generalized for other programming languages. Also, different programming languages might require different feature extraction methods due to their semantic differences. However, our data collections and analysis approaches could be easily generalized to packages written in any language. 

In addition, our dataset presented only open-source packages whose source code is hosted on GitHub that might not reflect close source packages. Also, in our study, we examined a dataset that contains 31 \npm JavaScript packages, which may not represent the whole population of JavaScript packages, and examining a larger number of packages may show different results.


	\section{Conclusion}
	\label{sec:conclusion}
	In this paper, our goal is to use ML techniques to determine \sv type of a new package release. We used 41 release-level features extracted by analyzing the source code and the development activities of the releases of 31 JavaScript packages published on \npm. Then, we built four ML classifiers.
We found that the XGBoost can effectively determine the type of \sv with average ROC-AUC equal to 0.77, 0.69, and 0.74 for major, minor, and patch releases. It also showed an improvement of 58\%, 38\%, and 49\% over our baseline, which is the ZeroR classifier. Regarding the most important features used by the XGBoost classifiers to determine \sv release type, we found that the change type and complexity and code dimensions of features are the most important indicators of new release type. Additionally, we investigated the generalizability of determining \sv type when we used cross-packages validation. Our results showed that the cross-packages validation achieves acceptable performance compared to within-packages validation.


				\bibliographystyle{cas-model2-names}
				
				\bibliography{bibliography}

\begin{thebibliography}{75}
\expandafter\ifx\csname natexlab\endcsname\relax\def\natexlab#1{#1}\fi
\providecommand{\url}[1]{\texttt{#1}}
\providecommand{\href}[2]{#2}
\providecommand{\path}[1]{#1}
\providecommand{\DOIprefix}{doi:}
\providecommand{\ArXivprefix}{arXiv:}
\providecommand{\URLprefix}{URL: }
\providecommand{\Pubmedprefix}{pmid:}
\providecommand{\doi}[1]{\href{http://dx.doi.org/#1}{\path{#1}}}
\providecommand{\Pubmed}[1]{\href{pmid:#1}{\path{#1}}}
\providecommand{\bibinfo}[2]{#2}
\ifx\xfnm\relax \def\xfnm[#1]{\unskip,\space#1}\fi
\bibitem[{{Abdalkareem} et~al.(2020){Abdalkareem}, {Mujahid} and
  {Shihab}}]{Abdalkareem2020TSE}
\bibinfo{author}{{Abdalkareem}, R.}, \bibinfo{author}{{Mujahid}, S.},
  \bibinfo{author}{{Shihab}, E.}, \bibinfo{year}{2020}.
\newblock \bibinfo{title}{A machine learning approach to improve the detection
  of ci skip commits}.
\newblock \bibinfo{journal}{IEEE Transactions on Software Engineering} ,
  \bibinfo{pages}{1--1}.
\bibitem[{Abdalkareem et~al.(2017)Abdalkareem, Nourry, Wehaibi, Mujahid and
  Shihab}]{AbdalkareemFSE2017}
\bibinfo{author}{Abdalkareem, R.}, \bibinfo{author}{Nourry, O.},
  \bibinfo{author}{Wehaibi, S.}, \bibinfo{author}{Mujahid, S.},
  \bibinfo{author}{Shihab, E.}, \bibinfo{year}{2017}.
\newblock \bibinfo{title}{Why do developers use trivial packages? an empirical
  case study on npm}, in: \bibinfo{booktitle}{Proceedings of the 2017 11th
  Joint Meeting on Foundations of Software Engineering},
  \bibinfo{publisher}{Association for Computing Machinery},
  \bibinfo{address}{New York, NY, USA}. p. \bibinfo{pages}{385–395}.
\newblock \URLprefix \url{https://doi.org/10.1145/3106237.3106267},
  \DOIprefix\doi{10.1145/3106237.3106267}.
\bibitem[{Ahasanuzzaman et~al.(2020)Ahasanuzzaman, Hassan and
  Hassan}]{ahasanuzzaman2020studying}
\bibinfo{author}{Ahasanuzzaman, M.}, \bibinfo{author}{Hassan, S.},
  \bibinfo{author}{Hassan, A.E.}, \bibinfo{year}{2020}.
\newblock \bibinfo{title}{Studying ad library integration strategies of top
  free-to-download apps}.
\newblock \bibinfo{journal}{IEEE Transactions on Software Engineering} .
\bibitem[{Alfassa(2013)}]{857922fo45online}
\bibinfo{author}{Alfassa, E.}, \bibinfo{year}{2013}.
\newblock \bibinfo{title}{857922 - fontconfig change breaks webfonts rendering
  under linux}.
\newblock
  \bibinfo{howpublished}{\url{https://bugzilla.mozilla.org/show_bug.cgi?id=857922}}.
\newblock \bibinfo{note}{(accessed on 02/25/2022)}.
\bibitem[{Andreasen et~al.(2017)Andreasen, Gong, M\o{}ller, Pradel, Selakovic,
  Sen and Staicu}]{AndreasenACM2017}
\bibinfo{author}{Andreasen, E.}, \bibinfo{author}{Gong, L.},
  \bibinfo{author}{M\o{}ller, A.}, \bibinfo{author}{Pradel, M.},
  \bibinfo{author}{Selakovic, M.}, \bibinfo{author}{Sen, K.},
  \bibinfo{author}{Staicu, C.A.}, \bibinfo{year}{2017}.
\newblock \bibinfo{title}{A survey of dynamic analysis and test generation for
  javascript}.
\newblock \bibinfo{journal}{ACM Comput. Surv.} \bibinfo{volume}{50}.
\bibitem[{Bacchelli et~al.(2012)Bacchelli, Dal~Sasso, D'Ambros and
  Lanza}]{BacchelliICSE2012}
\bibinfo{author}{Bacchelli, A.}, \bibinfo{author}{Dal~Sasso, T.},
  \bibinfo{author}{D'Ambros, M.}, \bibinfo{author}{Lanza, M.},
  \bibinfo{year}{2012}.
\newblock \bibinfo{title}{Content classification of development emails}, in:
  \bibinfo{booktitle}{Proceedings of the 34th International Conference on
  Software Engineering}, \bibinfo{publisher}{IEEE Press}. pp.
  \bibinfo{pages}{375--385}.
\bibitem[{{Bacchelli} et~al.(2012){Bacchelli}, {Dal Sasso}, {D'Ambros} and
  {Lanza}}]{Bacchelli2012ICSE}
\bibinfo{author}{{Bacchelli}, A.}, \bibinfo{author}{{Dal Sasso}, T.},
  \bibinfo{author}{{D'Ambros}, M.}, \bibinfo{author}{{Lanza}, M.},
  \bibinfo{year}{2012}.
\newblock \bibinfo{title}{Content classification of development emails}, in:
  \bibinfo{booktitle}{2012 34th International Conference on Software
  Engineering (ICSE)}, \bibinfo{publisher}{IEEE}. pp.
  \bibinfo{pages}{375--385}.
\bibitem[{Bengio and Grandvalet(2004)}]{bengio2004no}
\bibinfo{author}{Bengio, Y.}, \bibinfo{author}{Grandvalet, Y.},
  \bibinfo{year}{2004}.
\newblock \bibinfo{title}{No unbiased estimator of the variance of k-fold
  cross-validation}.
\newblock \bibinfo{journal}{Journal of machine learning research}
  \bibinfo{volume}{5}, \bibinfo{pages}{1089--1105}.
\bibitem[{Bogart et~al.(2017a)Bogart, Filippova, Kastner and
  Herbsleb}]{HowEcosy57online}
\bibinfo{author}{Bogart, C.}, \bibinfo{author}{Filippova, A.},
  \bibinfo{author}{Kastner, C.}, \bibinfo{author}{Herbsleb, J.},
  \bibinfo{year}{2017}a.
\newblock \bibinfo{title}{How ecosystem cultures differ: Results from a survey
  on values and practices across 18 software ecosystems}.
\newblock \bibinfo{howpublished}{\url{http://breakingapis.org/survey/}}.
\newblock \bibinfo{note}{(accessed on 11/17/2020)}.
\bibitem[{Bogart et~al.(2017b)Bogart, Filippova, Kästner and
  Herbsleb}]{ChristopherOnlineSurvey}
\bibinfo{author}{Bogart, C.}, \bibinfo{author}{Filippova, A.},
  \bibinfo{author}{Kästner, C.}, \bibinfo{author}{Herbsleb, J.},
  \bibinfo{year}{2017}b.
\newblock \bibinfo{title}{How ecosystem cultures differ: Results from a survey
  on values and practices across 18 software ecosystems}.
\newblock \bibinfo{howpublished}{[Online]. Available:
  http://breakingapis.org/survey/}.
\newblock \bibinfo{note}{(Accessed on 08/10/2020)}.
\bibitem[{Bogart et~al.(2016)Bogart, K\"{a}stner, Herbsleb and
  Thung}]{BogartFSE2016}
\bibinfo{author}{Bogart, C.}, \bibinfo{author}{K\"{a}stner, C.},
  \bibinfo{author}{Herbsleb, J.}, \bibinfo{author}{Thung, F.},
  \bibinfo{year}{2016}.
\newblock \bibinfo{title}{How to break an api: Cost negotiation and community
  values in three software ecosystems}, in: \bibinfo{booktitle}{Proceedings of
  the 2016 24th ACM SIGSOFT International Symposium on Foundations of Software
  Engineering}, \bibinfo{publisher}{Association for Computing Machinery},
  \bibinfo{address}{New York, NY, USA}. p. \bibinfo{pages}{109–120}.
\newblock \URLprefix \url{https://doi.org/10.1145/2950290.2950325},
  \DOIprefix\doi{10.1145/2950290.2950325}.
\bibitem[{Borges and Valente(2018)}]{BORGES2018112}
\bibinfo{author}{Borges, H.}, \bibinfo{author}{Valente, M.T.},
  \bibinfo{year}{2018}.
\newblock \bibinfo{title}{What’s in a github star? understanding repository
  starring practices in a social coding platform}.
\newblock \bibinfo{journal}{Journal of Systems and Software}
  \bibinfo{volume}{146}, \bibinfo{pages}{112 -- 129}.
\bibitem[{Bouckaert et~al.(2013)Bouckaert, Frank, Hall, Kirkby, Reutemann,
  Seewald and Scuse}]{WekaManu91online}
\bibinfo{author}{Bouckaert, R.R.}, \bibinfo{author}{Frank, E.},
  \bibinfo{author}{Hall, M.}, \bibinfo{author}{Kirkby, R.},
  \bibinfo{author}{Reutemann, P.}, \bibinfo{author}{Seewald, A.},
  \bibinfo{author}{Scuse, D.}, \bibinfo{year}{2013}.
\newblock \bibinfo{title}{WEKA Manual for Version 3-7-8}.
\newblock \bibinfo{note}{(accessed on 02/28/2021)}.
\bibitem[{Bradley(1997)}]{bradley1997use}
\bibinfo{author}{Bradley, A.P.}, \bibinfo{year}{1997}.
\newblock \bibinfo{title}{The use of the area under the roc curve in the
  evaluation of machine learning algorithms}.
\newblock \bibinfo{journal}{Pattern recognition} \bibinfo{volume}{30},
  \bibinfo{pages}{1145--1159}.
\bibitem[{Breiman(2001)}]{breiman2001random}
\bibinfo{author}{Breiman, L.}, \bibinfo{year}{2001}.
\newblock \bibinfo{title}{Random forests}.
\newblock \bibinfo{journal}{Machine learning} \bibinfo{volume}{45},
  \bibinfo{pages}{5--32}.
\bibitem[{Businge et~al.(2012)Businge, Serebrenik and van~den
  Brand}]{Businge_ICSM2012}
\bibinfo{author}{Businge, J.}, \bibinfo{author}{Serebrenik, A.},
  \bibinfo{author}{van~den Brand, M.G.J.}, \bibinfo{year}{2012}.
\newblock \bibinfo{title}{Survival of eclipse third-party plug-ins}, in:
  \bibinfo{booktitle}{Proceedings of the 28th IEEE International Conference on
  Software Maintenance}, \bibinfo{publisher}{IEEE}, \bibinfo{address}{New York,
  NY, USA}. pp. \bibinfo{pages}{368--377}.
\newblock \DOIprefix\doi{10.1109/ICSM.2012.6405295}.
\bibitem[{Businge et~al.(2015)Businge, Serebrenik and van~den
  Brand}]{Businge_SQJ2015}
\bibinfo{author}{Businge, J.}, \bibinfo{author}{Serebrenik, A.},
  \bibinfo{author}{van~den Brand, M.G.J.}, \bibinfo{year}{2015}.
\newblock \bibinfo{title}{Eclipse api usage: The good and the bad}.
\newblock \bibinfo{journal}{Software Quality Journal} \bibinfo{volume}{23},
  \bibinfo{pages}{107–141}.
\newblock \DOIprefix\doi{10.1007/s11219-013-9221-3}.
\bibitem[{Caruana and Niculescu-Mizil(2006)}]{CaruanaICML2006}
\bibinfo{author}{Caruana, R.}, \bibinfo{author}{Niculescu-Mizil, A.},
  \bibinfo{year}{2006}.
\newblock \bibinfo{title}{An empirical comparison of supervised learning
  algorithms}, in: \bibinfo{booktitle}{Proceedings of the 23rd International
  Conference on Machine Learning}, \bibinfo{publisher}{ACM}. pp.
  \bibinfo{pages}{161--168}.
\bibitem[{Castelluccio et~al.(2019)Castelluccio, An and
  Khomh}]{castelluccio2019empirical}
\bibinfo{author}{Castelluccio, M.}, \bibinfo{author}{An, L.},
  \bibinfo{author}{Khomh, F.}, \bibinfo{year}{2019}.
\newblock \bibinfo{title}{An empirical study of patch uplift in rapid release
  development pipelines}.
\newblock \bibinfo{journal}{Empirical Software Engineering}
  \bibinfo{volume}{24}, \bibinfo{pages}{3008--3044}.
\bibitem[{Chawla et~al.(2002)Chawla, Bowyer, Hall and
  Kegelmeyer}]{chawla2002smote}
\bibinfo{author}{Chawla, N.V.}, \bibinfo{author}{Bowyer, K.W.},
  \bibinfo{author}{Hall, L.O.}, \bibinfo{author}{Kegelmeyer, W.P.},
  \bibinfo{year}{2002}.
\newblock \bibinfo{title}{Smote: synthetic minority over-sampling technique}.
\newblock \bibinfo{journal}{Journal of artificial intelligence research}
  \bibinfo{volume}{16}, \bibinfo{pages}{321--357}.
\bibitem[{Chen and Guestrin(2016)}]{TianqiKDD2016}
\bibinfo{author}{Chen, T.}, \bibinfo{author}{Guestrin, C.},
  \bibinfo{year}{2016}.
\newblock \bibinfo{title}{Xgboost: A scalable tree boosting system}, in:
  \bibinfo{booktitle}{Proceedings of the 22nd ACM SIGKDD International
  Conference on Knowledge Discovery and Data Mining},
  \bibinfo{publisher}{Association for Computing Machinery},
  \bibinfo{address}{New York, NY, USA}. p. \bibinfo{pages}{785–794}.
\newblock \URLprefix \url{https://doi.org/10.1145/2939672.2939785},
  \DOIprefix\doi{10.1145/2939672.2939785}.
\bibitem[{Dabbish et~al.(2012)Dabbish, Stuart, Tsay and
  Herbsleb}]{DabbishCSCW2012}
\bibinfo{author}{Dabbish, L.}, \bibinfo{author}{Stuart, C.},
  \bibinfo{author}{Tsay, J.}, \bibinfo{author}{Herbsleb, J.},
  \bibinfo{year}{2012}.
\newblock \bibinfo{title}{Social coding in github: Transparency and
  collaboration in an open software repository}, in:
  \bibinfo{booktitle}{Proceedings of the ACM 2012 Conference on Computer
  Supported Cooperative Work}, \bibinfo{publisher}{ACM}. pp.
  \bibinfo{pages}{1277--1286}.
\bibitem[{{Decan} and {Mens}(2019)}]{DecanTSE2019}
\bibinfo{author}{{Decan}, A.}, \bibinfo{author}{{Mens}, T.},
  \bibinfo{year}{2019}.
\newblock \bibinfo{title}{What do package dependencies tell us about semantic
  versioning?}
\newblock \bibinfo{journal}{IEEE Transactions on Software Engineering} ,
  \bibinfo{pages}{1--15}.
\bibitem[{Decan et~al.(2019)Decan, Mens and Grosjean}]{DecanEMSE2019}
\bibinfo{author}{Decan, A.}, \bibinfo{author}{Mens, T.},
  \bibinfo{author}{Grosjean, P.}, \bibinfo{year}{2019}.
\newblock \bibinfo{title}{An empirical comparison of dependency network
  evolution in seven software packaging ecosystems}.
\newblock \bibinfo{journal}{Empirical Software Engineering} ,
  \bibinfo{pages}{381--416}.
\bibitem[{{Dietrich} et~al.(2019){Dietrich}, {Pearce}, {Stringer}, {Tahir} and
  {Blincoe}}]{DietrichMSR2019}
\bibinfo{author}{{Dietrich}, J.}, \bibinfo{author}{{Pearce}, D.},
  \bibinfo{author}{{Stringer}, J.}, \bibinfo{author}{{Tahir}, A.},
  \bibinfo{author}{{Blincoe}, K.}, \bibinfo{year}{2019}.
\newblock \bibinfo{title}{Dependency versioning in the wild}, in:
  \bibinfo{booktitle}{2019 IEEE/ACM 16th International Conference on Mining
  Software Repositories (MSR)}, pp. \bibinfo{pages}{349--359}.
\newblock \DOIprefix\doi{10.1109/MSR.2019.00061}.
\bibitem[{Dig and Johnson(2006)}]{Dig2006AES}
\bibinfo{author}{Dig, D.}, \bibinfo{author}{Johnson, R.}, \bibinfo{year}{2006}.
\newblock \bibinfo{title}{How do apis evolve\&quest; a story of refactoring}.
\newblock \bibinfo{journal}{Journal of Software Maintenance}
  \bibinfo{volume}{18}, \bibinfo{pages}{83--107}.
\newblock \DOIprefix\doi{10.1002/smr.328}.
\bibitem[{npm documentation()}]{AboutsSVonline}
\bibinfo{author}{npm documentation}, .
\newblock \bibinfo{title}{About semantic versioning | npm docs}.
\newblock
  \bibinfo{howpublished}{\url{https://docs.npmjs.com/about-semantic-versioning}}.
\newblock \bibinfo{note}{(accessed on 03/08/2022)}.
\bibitem[{Esteves et~al.(2020)Esteves, Figueiredo, Veloso, Viggiato and
  Ziviani}]{esteves2020understanding}
\bibinfo{author}{Esteves, G.}, \bibinfo{author}{Figueiredo, E.},
  \bibinfo{author}{Veloso, A.}, \bibinfo{author}{Viggiato, M.},
  \bibinfo{author}{Ziviani, N.}, \bibinfo{year}{2020}.
\newblock \bibinfo{title}{Understanding machine learning software defect
  predictions}.
\newblock \bibinfo{journal}{Automated Software Engineering}
  \bibinfo{volume}{27}, \bibinfo{pages}{369--392}.
\bibitem[{FaceBook(2016)}]{YarnAnew90online}
\bibinfo{author}{FaceBook}, \bibinfo{year}{2016}.
\newblock \bibinfo{title}{Yarn: A new package manager for javascript - facebook
  engineering}.
\newblock
  \bibinfo{howpublished}{\url{https://engineering.fb.com/2016/10/11/web/yarn-a-new-package-manager-for-javascript/}}.
\newblock \bibinfo{note}{(accessed on 03/13/2021)}.
\bibitem[{Falleri et~al.(2014)Falleri, Morandat, Blanc, Martinez and
  Monperrus}]{FalleriGumTreeASE2014}
\bibinfo{author}{Falleri, J.}, \bibinfo{author}{Morandat, F.},
  \bibinfo{author}{Blanc, X.}, \bibinfo{author}{Martinez, M.},
  \bibinfo{author}{Monperrus, M.}, \bibinfo{year}{2014}.
\newblock \bibinfo{title}{Fine-grained and accurate source code differencing},
  in: \bibinfo{booktitle}{{ACM/IEEE} International Conference on Automated
  Software Engineering, {ASE} '14, Vasteras, Sweden - September 15 - 19, 2014},
  pp. \bibinfo{pages}{313--324}.
\newblock \URLprefix \url{http://doi.acm.org/10.1145/2642937.2642982},
  \DOIprefix\doi{10.1145/2642937.2642982}.
\bibitem[{Fukushima et~al.(2014)Fukushima, Kamei, McIntosh, Yamashita and
  Ubayashi}]{Takafumi2014MSR}
\bibinfo{author}{Fukushima, T.}, \bibinfo{author}{Kamei, Y.},
  \bibinfo{author}{McIntosh, S.}, \bibinfo{author}{Yamashita, K.},
  \bibinfo{author}{Ubayashi, N.}, \bibinfo{year}{2014}.
\newblock \bibinfo{title}{An empirical study of just-in-time defect prediction
  using cross-project models}, in: \bibinfo{booktitle}{Proceedings of the 11th
  Working Conference on Mining Software Repositories},
  \bibinfo{publisher}{Association for Computing Machinery}. p.
  \bibinfo{pages}{172–181}.
\bibitem[{Ghotra et~al.(2015)Ghotra, ~ and Hassan}]{GhotraICSE2015}
\bibinfo{author}{Ghotra, B.}, \bibinfo{author}{~, S.}, \bibinfo{author}{Hassan,
  A.E.}, \bibinfo{year}{2015}.
\newblock \bibinfo{title}{Revisiting the impact of classification techniques on
  the performance of defect prediction models}, in:
  \bibinfo{booktitle}{Proceedings of the 37th International Conference on
  Software Engineering}, \bibinfo{publisher}{IEEE Press}. pp.
  \bibinfo{pages}{789--800}.
\bibitem[{Grissom and Kim(2005)}]{grissom2005effect}
\bibinfo{author}{Grissom, R.J.}, \bibinfo{author}{Kim, J.J.},
  \bibinfo{year}{2005}.
\newblock \bibinfo{title}{Effect sizes for research: A broad practical
  approach.}
\newblock \bibinfo{publisher}{Lawrence Erlbaum Associates Publishers}.
\bibitem[{Hall et~al.(2009)Hall, Frank, Holmes, Pfahringer, Reutemann and
  Witten}]{hall2009weka}
\bibinfo{author}{Hall, M.}, \bibinfo{author}{Frank, E.},
  \bibinfo{author}{Holmes, G.}, \bibinfo{author}{Pfahringer, B.},
  \bibinfo{author}{Reutemann, P.}, \bibinfo{author}{Witten, I.H.},
  \bibinfo{year}{2009}.
\newblock \bibinfo{title}{The weka data mining software: an update}.
\newblock \bibinfo{journal}{ACM SIGKDD explorations newsletter}
  \bibinfo{volume}{11}, \bibinfo{pages}{10--18}.
\bibitem[{He et~al.(2012)He, Shu, Yang, Li and Wang}]{HeASEJ2012}
\bibinfo{author}{He, Z.}, \bibinfo{author}{Shu, F.}, \bibinfo{author}{Yang,
  Y.}, \bibinfo{author}{Li, M.}, \bibinfo{author}{Wang, Q.},
  \bibinfo{year}{2012}.
\newblock \bibinfo{title}{An investigation on the feasibility of cross-project
  defect prediction}.
\newblock \bibinfo{journal}{Automated Software Engineering.}
  \bibinfo{volume}{19}, \bibinfo{pages}{167--199}.
\bibitem[{Iba(1996)}]{IbaPPSN1996}
\bibinfo{author}{Iba, H.}, \bibinfo{year}{1996}.
\newblock \bibinfo{title}{Random tree generation for genetic programming}, in:
  \bibinfo{booktitle}{Proceedings of the 4th International Conference on
  Parallel Problem Solving from Nature}, \bibinfo{publisher}{Springer-Verlag},
  \bibinfo{address}{London, UK, UK}. pp. \bibinfo{pages}{144--153}.
\newblock \URLprefix \url{http://dl.acm.org/citation.cfm?id=645823.670546}.
\bibitem[{Javan~Jafari et~al.(2021)Javan~Jafari, Costa, Abdalkareem, Shihab and
  Tsantalis}]{AbbasTSE2021}
\bibinfo{author}{Javan~Jafari, A.}, \bibinfo{author}{Costa, D.E.},
  \bibinfo{author}{Abdalkareem, R.}, \bibinfo{author}{Shihab, E.},
  \bibinfo{author}{Tsantalis, N.}, \bibinfo{year}{2021}.
\newblock \bibinfo{title}{Dependency smells in javascript projects}.
\newblock \bibinfo{journal}{IEEE Transactions on Software Engineering} ,
  \bibinfo{pages}{1--1}\DOIprefix\doi{10.1109/TSE.2021.3106247}.
\bibitem[{Kamei et~al.(2013)Kamei, Shihab, Adams, Hassan, Mockus, Sinha and
  Ubayashi}]{kameiTSE2013}
\bibinfo{author}{Kamei, Y.}, \bibinfo{author}{Shihab, E.},
  \bibinfo{author}{Adams, B.}, \bibinfo{author}{Hassan, A.E.},
  \bibinfo{author}{Mockus, A.}, \bibinfo{author}{Sinha, A.},
  \bibinfo{author}{Ubayashi, N.}, \bibinfo{year}{2013}.
\newblock \bibinfo{title}{A large-scale empirical study of just-in-time quality
  assurance}.
\newblock \bibinfo{journal}{IEEE Transactions on Software Engineering}
  \bibinfo{volume}{39}, \bibinfo{pages}{757--773}.
\bibitem[{Kapur et~al.(2010)Kapur, Cossette and Walker}]{Kapur2010RR}
\bibinfo{author}{Kapur, P.}, \bibinfo{author}{Cossette, B.},
  \bibinfo{author}{Walker, R.J.}, \bibinfo{year}{2010}.
\newblock \bibinfo{title}{Refactoring references for library migration}.
\newblock \bibinfo{journal}{ACM SIGPLAN Notices} \bibinfo{volume}{45},
  \bibinfo{pages}{726--738}.
\newblock \DOIprefix\doi{10.1145/1932682.1869518}.
\bibitem[{Khomh et~al.(2015)Khomh, Adams, Dhaliwal and Zou}]{KhomhEMSE2015}
\bibinfo{author}{Khomh, F.}, \bibinfo{author}{Adams, B.},
  \bibinfo{author}{Dhaliwal, T.}, \bibinfo{author}{Zou, Y.},
  \bibinfo{year}{2015}.
\newblock \bibinfo{title}{Understanding the impact of rapid releases on
  software quality}.
\newblock \bibinfo{journal}{Empirical Softw. Engg.} \bibinfo{volume}{20},
  \bibinfo{pages}{336–373}.
\newblock \URLprefix \url{https://doi.org/10.1007/s10664-014-9308-x},
  \DOIprefix\doi{10.1007/s10664-014-9308-x}.
\bibitem[{Kotsiantis et~al.(2006)Kotsiantis, Zaharakis and
  Pintelas}]{Kotsiantis2006}
\bibinfo{author}{Kotsiantis, S.B.}, \bibinfo{author}{Zaharakis, I.D.},
  \bibinfo{author}{Pintelas, P.E.}, \bibinfo{year}{2006}.
\newblock \bibinfo{title}{Machine learning: A review of classification and
  combining techniques}.
\newblock \bibinfo{journal}{Artif. Intell. Rev.} \bibinfo{volume}{26},
  \bibinfo{pages}{159--190}.
\bibitem[{Kula et~al.(2017)Kula, German, Ouni, Ishio and Inoue}]{Kula_EMSE2017}
\bibinfo{author}{Kula, R.G.}, \bibinfo{author}{German, D.M.},
  \bibinfo{author}{Ouni, A.}, \bibinfo{author}{Ishio, T.},
  \bibinfo{author}{Inoue, K.}, \bibinfo{year}{2017}.
\newblock \bibinfo{title}{{Do developers update their library dependencies?: An
  empirical study on the impact of security advisories on library migration}}.
\newblock \DOIprefix\doi{10.1007/s10664-017-9521-5},
  \href{http://arxiv.org/abs/1709.04621}{\tt arXiv:1709.04621}.
\bibitem[{Lauinger et~al.(2018)Lauinger, Chaabane and
  Wilson}]{LauingerDependency}
\bibinfo{author}{Lauinger, T.}, \bibinfo{author}{Chaabane, A.},
  \bibinfo{author}{Wilson, C.}, \bibinfo{year}{2018}.
\newblock \bibinfo{title}{Thou shalt not depend on me: A look at javascript
  libraries in the wild}.
\newblock \bibinfo{journal}{Queue} \bibinfo{volume}{16},
  \bibinfo{pages}{62–82}.
\bibitem[{{Lessmann} et~al.(2008){Lessmann}, {Baesens}, {Mues} and
  {Pietsch}}]{LessmannTSE2008}
\bibinfo{author}{{Lessmann}, S.}, \bibinfo{author}{{Baesens}, B.},
  \bibinfo{author}{{Mues}, C.}, \bibinfo{author}{{Pietsch}, S.},
  \bibinfo{year}{2008}.
\newblock \bibinfo{title}{Benchmarking classification models for software
  defect prediction: A proposed framework and novel findings}.
\newblock \bibinfo{journal}{IEEE Transactions on Software Engineering}
  \bibinfo{volume}{34}, \bibinfo{pages}{485--496}.
\bibitem[{Mann and Whitney(1947)}]{mann1947test}
\bibinfo{author}{Mann, H.B.}, \bibinfo{author}{Whitney, D.R.},
  \bibinfo{year}{1947}.
\newblock \bibinfo{title}{On a test of whether one of two random variables is
  stochastically larger than the other}.
\newblock \bibinfo{journal}{The annals of mathematical statistics} ,
  \bibinfo{pages}{50--60}.
\bibitem[{{Mariano} et~al.(2019){Mariano}, {dos Santos}, {V. de Almeida} and
  {Brandão}}]{Mariano2019ICMLA}
\bibinfo{author}{{Mariano}, R.V.R.}, \bibinfo{author}{{dos Santos}, G.E.},
  \bibinfo{author}{{V. de Almeida}, M.}, \bibinfo{author}{{Brandão}, W.C.},
  \bibinfo{year}{2019}.
\newblock \bibinfo{title}{Feature changes in source code for commit
  classification into maintenance activities}, in: \bibinfo{booktitle}{2019
  18th IEEE International Conference On Machine Learning And Applications
  (ICMLA)}, \bibinfo{publisher}{IEEE}. pp. \bibinfo{pages}{515--518}.
\bibitem[{Mostafa et~al.(2017a)Mostafa, Rodriguez and Wang}]{Mostafa2017}
\bibinfo{author}{Mostafa, S.}, \bibinfo{author}{Rodriguez, R.},
  \bibinfo{author}{Wang, X.}, \bibinfo{year}{2017}a.
\newblock \bibinfo{title}{{A Study on Behavioral Backward Incompatibility Bugs
  in Java Software Libraries}}, in: \bibinfo{booktitle}{Proceedings of the 39th
  International Conference on Software Engineering Companion},
  \bibinfo{publisher}{IEEE}, \bibinfo{address}{New York, NY, USA}. pp.
  \bibinfo{pages}{127--129}.
\newblock \DOIprefix\doi{10.1109/ICSE-C.2017.101}.
\bibitem[{Mostafa et~al.(2017b)Mostafa, Rodriguez and Wang}]{MostafaISSTA2017}
\bibinfo{author}{Mostafa, S.}, \bibinfo{author}{Rodriguez, R.},
  \bibinfo{author}{Wang, X.}, \bibinfo{year}{2017}b.
\newblock \bibinfo{title}{Experience paper: A study on behavioral backward
  incompatibilities of java software libraries}, in:
  \bibinfo{booktitle}{Proceedings of the 26th ACM SIGSOFT International
  Symposium on Software Testing and Analysis}, \bibinfo{publisher}{Association
  for Computing Machinery}, \bibinfo{address}{New York, NY, USA}. p.
  \bibinfo{pages}{215–225}.
\newblock \URLprefix \url{https://doi.org/10.1145/3092703.3092721},
  \DOIprefix\doi{10.1145/3092703.3092721}.
\bibitem[{Mujahid et~al.(2020)Mujahid, Abdalkareem, Shihab and
  McIntosh}]{mujahidusingMSR2020}
\bibinfo{author}{Mujahid, S.}, \bibinfo{author}{Abdalkareem, R.},
  \bibinfo{author}{Shihab, E.}, \bibinfo{author}{McIntosh, S.},
  \bibinfo{year}{2020}.
\newblock \bibinfo{title}{Using others’ tests to identify breaking updates} ,
  \bibinfo{pages}{1--12}.
\bibitem[{Murphy(2012)}]{murphy2012machine}
\bibinfo{author}{Murphy, K.P.}, \bibinfo{year}{2012}.
\newblock \bibinfo{title}{Machine learning: a probabilistic perspective}.
\newblock \bibinfo{publisher}{MIT press}.
\bibitem[{Nam and Kim(2015)}]{NamASE2015}
\bibinfo{author}{Nam, J.}, \bibinfo{author}{Kim, S.}, \bibinfo{year}{2015}.
\newblock \bibinfo{title}{Clami: Defect prediction on unlabeled datasets}, in:
  \bibinfo{booktitle}{Proceedings of the 30th IEEE/ACM International Conference
  on Automated Software Engineering}, \bibinfo{publisher}{IEEE Press}. p.
  \bibinfo{pages}{452–463}.
\bibitem[{npm()}]{npmregis30online}
\bibinfo{author}{npm}, .
\newblock \bibinfo{title}{npm-registry | npm documentation}.
\newblock
  \bibinfo{howpublished}{\url{https://docs.npmjs.com/using-npm/registry.html}}.
\newblock \bibinfo{note}{(Accessed on 08/13/2020)}.
\bibitem[{Pedregosa et~al.(2011)Pedregosa, Varoquaux, Gramfort, Michel,
  Thirion, Grisel, Blondel, Prettenhofer, Weiss, Dubourg
  et~al.}]{pedregosa2011scikit}
\bibinfo{author}{Pedregosa, F.}, \bibinfo{author}{Varoquaux, G.},
  \bibinfo{author}{Gramfort, A.}, \bibinfo{author}{Michel, V.},
  \bibinfo{author}{Thirion, B.}, \bibinfo{author}{Grisel, O.},
  \bibinfo{author}{Blondel, M.}, \bibinfo{author}{Prettenhofer, P.},
  \bibinfo{author}{Weiss, R.}, \bibinfo{author}{Dubourg, V.}, et~al.,
  \bibinfo{year}{2011}.
\newblock \bibinfo{title}{Scikit-learn: Machine learning in python}.
\newblock \bibinfo{journal}{the Journal of machine Learning research}
  \bibinfo{volume}{12}, \bibinfo{pages}{2825--2830}.
\bibitem[{Potvin and Levenberg(2016)}]{potvin2016google}
\bibinfo{author}{Potvin, R.}, \bibinfo{author}{Levenberg, J.},
  \bibinfo{year}{2016}.
\newblock \bibinfo{title}{Why google stores billions of lines of code in a
  single repository}.
\newblock \bibinfo{journal}{Communications of the ACM} \bibinfo{volume}{59},
  \bibinfo{pages}{78--87}.
\bibitem[{Pradel et~al.(2015)Pradel, Schuh and Sen}]{pradel2015typedevil}
\bibinfo{author}{Pradel, M.}, \bibinfo{author}{Schuh, P.},
  \bibinfo{author}{Sen, K.}, \bibinfo{year}{2015}.
\newblock \bibinfo{title}{Typedevil: Dynamic type inconsistency analysis for
  javascript}, in: \bibinfo{booktitle}{2015 IEEE/ACM 37th IEEE International
  Conference on Software Engineering}, \bibinfo{organization}{IEEE}. pp.
  \bibinfo{pages}{314--324}.
\bibitem[{Preston-Werner(2019)}]{semver2019}
\bibinfo{author}{Preston-Werner, T.}, \bibinfo{year}{2019}.
\newblock \bibinfo{title}{Semantic versioning 2.0}.
\newblock \URLprefix \url{https://semver.org/}.
\bibitem[{Quinlan(1993)}]{Quinlan1993}
\bibinfo{author}{Quinlan, R.}, \bibinfo{year}{1993}.
\newblock \bibinfo{title}{C4.5: Programs for Machine Learning}.
\newblock \bibinfo{publisher}{Morgan Kaufmann Publishers},
  \bibinfo{address}{San Mateo, CA}.
\bibitem[{Raemaekers et~al.(2017)Raemaekers, van Deursen and
  Visser}]{raemaekers2017semantic}
\bibinfo{author}{Raemaekers, S.}, \bibinfo{author}{van Deursen, A.},
  \bibinfo{author}{Visser, J.}, \bibinfo{year}{2017}.
\newblock \bibinfo{title}{Semantic versioning and impact of breaking changes in
  the maven repository}.
\newblock \bibinfo{journal}{Journal of Systems and Software}
  \bibinfo{volume}{129}, \bibinfo{pages}{140--158}.
\bibitem[{Rahman et~al.(2017)Rahman, Roy and Kula}]{RahmanMSR2017}
\bibinfo{author}{Rahman, M.M.}, \bibinfo{author}{Roy, C.K.},
  \bibinfo{author}{Kula, R.G.}, \bibinfo{year}{2017}.
\newblock \bibinfo{title}{Predicting usefulness of code review comments using
  textual features and developer experience}, in:
  \bibinfo{booktitle}{Proceedings of the 14th International Conference on
  Mining Software Repositories}, \bibinfo{publisher}{IEEE Press}. pp.
  \bibinfo{pages}{215--226}.
\bibitem[{Rahman et~al.(2019)Rahman, Rigby and Shihab}]{rahman2019modular}
\bibinfo{author}{Rahman, M.T.}, \bibinfo{author}{Rigby, P.C.},
  \bibinfo{author}{Shihab, E.}, \bibinfo{year}{2019}.
\newblock \bibinfo{title}{The modular and feature toggle architectures of
  google chrome}.
\newblock \bibinfo{journal}{Empirical Software Engineering}
  \bibinfo{volume}{24}, \bibinfo{pages}{826--853}.
\bibitem[{Reza~Chowdhury et~al.(2021)Reza~Chowdhury, Abdalkareem, Shihab and
  Adams}]{ChowdhuryTSE2021}
\bibinfo{author}{Reza~Chowdhury, M.A.}, \bibinfo{author}{Abdalkareem, R.},
  \bibinfo{author}{Shihab, E.}, \bibinfo{author}{Adams, B.},
  \bibinfo{year}{2021}.
\newblock \bibinfo{title}{On the untriviality of trivial packages: An empirical
  study of npm javascript packages}.
\newblock \bibinfo{journal}{IEEE Transactions on Software Engineering} ,
  \bibinfo{pages}{1--1}.
\bibitem[{SciTools-Documentation()}]{SciTools2Online}
\bibinfo{author}{SciTools-Documentation}, .
\newblock \bibinfo{title}{Understand static code analysis tool}.
\newblock \bibinfo{howpublished}{\url{https://www.scitools.com/}}.
\newblock \bibinfo{note}{(accessed on 03/08/2022)}.
\bibitem[{Shihab et~al.(2010)Shihab, Jiang, Ibrahim, Adams and
  Hassan}]{ShihabESEM2010}
\bibinfo{author}{Shihab, E.}, \bibinfo{author}{Jiang, Z.M.},
  \bibinfo{author}{Ibrahim, W.M.}, \bibinfo{author}{Adams, B.},
  \bibinfo{author}{Hassan, A.E.}, \bibinfo{year}{2010}.
\newblock \bibinfo{title}{Understanding the impact of code and process metrics
  on post-release defects: A case study on the eclipse project}, in:
  \bibinfo{booktitle}{Proceedings of the 2010 ACM-IEEE International Symposium
  on Empirical Software Engineering and Measurement}.
\bibitem[{{\'S}liwerski et~al.(2005){\'S}liwerski, Zimmermann and
  Zeller}]{sliwerski2005changes}
\bibinfo{author}{{\'S}liwerski, J.}, \bibinfo{author}{Zimmermann, T.},
  \bibinfo{author}{Zeller, A.}, \bibinfo{year}{2005}.
\newblock \bibinfo{title}{When do changes induce fixes?}
\newblock \bibinfo{journal}{ACM sigsoft software engineering notes}
  \bibinfo{volume}{30}, \bibinfo{pages}{1--5}.
\bibitem[{{Song} et~al.(2019){Song}, {Guo} and {Shepperd}}]{SongTSE2019}
\bibinfo{author}{{Song}, Q.}, \bibinfo{author}{{Guo}, Y.},
  \bibinfo{author}{{Shepperd}, M.}, \bibinfo{year}{2019}.
\newblock \bibinfo{title}{A comprehensive investigation of the role of
  imbalanced learning for software defect prediction}.
\newblock \bibinfo{journal}{IEEE Transactions on Software Engineering}
  \bibinfo{volume}{45}, \bibinfo{pages}{1253--1269}.
\newblock \DOIprefix\doi{10.1109/TSE.2018.2836442}.
\bibitem[{Souza et~al.(2014)Souza, Chavez and Bittencourt}]{Souza}
\bibinfo{author}{Souza, R.}, \bibinfo{author}{Chavez, C.},
  \bibinfo{author}{Bittencourt, R.A.}, \bibinfo{year}{2014}.
\newblock \bibinfo{title}{Do rapid releases affect bug reopening? a case study
  of firefox}, in: \bibinfo{booktitle}{2014 Brazilian Symposium on Software
  Engineering}, pp. \bibinfo{pages}{31--40}.
\newblock \DOIprefix\doi{10.1109/SBES.2014.10}.
\bibitem[{Thung et~al.(2012)Thung, Lo, Jiang, Lucia, Rahman and
  Devanbu}]{ThungICSM2012}
\bibinfo{author}{Thung, F.}, \bibinfo{author}{Lo, D.}, \bibinfo{author}{Jiang,
  L.}, \bibinfo{author}{Lucia}, \bibinfo{author}{Rahman, F.},
  \bibinfo{author}{Devanbu, P.T.}, \bibinfo{year}{2012}.
\newblock \bibinfo{title}{When would this bug get reported?}, in:
  \bibinfo{booktitle}{Proceedings of the 28th IEEE International Conference on
  Software Maintenance}, \bibinfo{publisher}{IEEE}. pp.
  \bibinfo{pages}{420--429}.
\bibitem[{Understand()}]{SciTools70online}
\bibinfo{author}{Understand, S.}, .
\newblock \bibinfo{title}{Scitools.com}.
\newblock \bibinfo{howpublished}{\url{https://scitools.com/}}.
\newblock \bibinfo{note}{(Accessed on 08/13/2020)}.
\bibitem[{Williams and Spacco(2008)}]{williams2008szz}
\bibinfo{author}{Williams, C.}, \bibinfo{author}{Spacco, J.},
  \bibinfo{year}{2008}.
\newblock \bibinfo{title}{Szz revisited: verifying when changes induce fixes},
  in: \bibinfo{booktitle}{Proceedings of the 2008 workshop on Defects in large
  software systems}, pp. \bibinfo{pages}{32--36}.
\bibitem[{Wittern et~al.(2016)Wittern, Suter and Rajagopalan}]{WitternMSR2016}
\bibinfo{author}{Wittern, E.}, \bibinfo{author}{Suter, P.},
  \bibinfo{author}{Rajagopalan, S.}, \bibinfo{year}{2016}.
\newblock \bibinfo{title}{A look at the dynamics of the javascript package
  ecosystem}, in: \bibinfo{booktitle}{Proceedings of the 13th International
  Conference on Mining Software Repositories}, \bibinfo{publisher}{Association
  for Computing Machinery}, \bibinfo{address}{New York, NY, USA}. p.
  \bibinfo{pages}{351–361}.
\newblock \URLprefix \url{https://doi.org/10.1145/2901739.2901743},
  \DOIprefix\doi{10.1145/2901739.2901743}.
\bibitem[{{Xavier} et~al.(2017){Xavier}, {Brito}, {Hora} and
  {Valente}}]{XavierSANER2017}
\bibinfo{author}{{Xavier}, L.}, \bibinfo{author}{{Brito}, A.},
  \bibinfo{author}{{Hora}, A.}, \bibinfo{author}{{Valente}, M.T.},
  \bibinfo{year}{2017}.
\newblock \bibinfo{title}{Historical and impact analysis of api breaking
  changes: A large-scale study}, in: \bibinfo{booktitle}{2017 IEEE 24th
  International Conference on Software Analysis, Evolution and Reengineering
  (SANER)}, \bibinfo{publisher}{IEEE}. pp. \bibinfo{pages}{138--147}.
\bibitem[{Xavier et~al.(2017)Xavier, Brito, Hora and
  Valente}]{Xavier_SANER2017}
\bibinfo{author}{Xavier, L.}, \bibinfo{author}{Brito, A.},
  \bibinfo{author}{Hora, A.}, \bibinfo{author}{Valente, M.T.},
  \bibinfo{year}{2017}.
\newblock \bibinfo{title}{Historical and impact analysis of api breaking
  changes: A large-scale study}, in: \bibinfo{booktitle}{Proceedings of the
  24th International Conference on Software Analysis, Evolution and
  Reengineering}, \bibinfo{publisher}{IEEE}, \bibinfo{address}{New York, NY,
  USA}. pp. \bibinfo{pages}{138--147}.
\newblock \DOIprefix\doi{10.1109/SANER.2017.7884616}.
\bibitem[{Xia et~al.(2016a)Xia, ~, Kamei, Lo and Wang}]{XiaESEM2016}
\bibinfo{author}{Xia, X.}, \bibinfo{author}{~, E.}, \bibinfo{author}{Kamei,
  Y.}, \bibinfo{author}{Lo, D.}, \bibinfo{author}{Wang, X.},
  \bibinfo{year}{2016}a.
\newblock \bibinfo{title}{Predicting crashing releases of mobile applications},
  in: \bibinfo{booktitle}{Proceedings of the 10th ACM/IEEE International
  Symposium on Empirical Software Engineering and Measurement},
  \bibinfo{publisher}{ACM}. pp. \bibinfo{pages}{29:1--29:10}.
\bibitem[{Xia et~al.(2016b)Xia, Shihab, Kamei, Lo and Wang}]{Xia_ESEM2016}
\bibinfo{author}{Xia, X.}, \bibinfo{author}{Shihab, E.},
  \bibinfo{author}{Kamei, Y.}, \bibinfo{author}{Lo, D.}, \bibinfo{author}{Wang,
  X.}, \bibinfo{year}{2016}b.
\newblock \bibinfo{title}{Predicting crashing releases of mobile applications},
  in: \bibinfo{booktitle}{Proceedings of the 10th ACM/IEEE International
  Symposium on Empirical Software Engineering and Measurement (ESEM'16)}.
\bibitem[{{Yan} et~al.(2019){Yan}, {Xia}, {Shihab}, {Lo}, {Yin} and
  {Yang}}]{YanTSE2019}
\bibinfo{author}{{Yan}, M.}, \bibinfo{author}{{Xia}, X.},
  \bibinfo{author}{{Shihab}, E.}, \bibinfo{author}{{Lo}, D.},
  \bibinfo{author}{{Yin}, J.}, \bibinfo{author}{{Yang}, X.},
  \bibinfo{year}{2019}.
\newblock \bibinfo{title}{Automating change-level self-admitted technical debt
  determination}.
\newblock \bibinfo{journal}{IEEE Transactions on Software Engineering}
  \bibinfo{volume}{45}, \bibinfo{pages}{1211--1229}.

\end{thebibliography}

%
%
%
				
			\end{document}